\documentclass[fleqn,usenatbib]{mnras}
\usepackage{newtxtext,newtxmath}

\usepackage[T1]{fontenc}

\DeclareRobustCommand{\VAN}[3]{#2}
\let\VANthebibliography\thebibliography
\def\thebibliography{\DeclareRobustCommand{\VAN}[3]{##3}\VANthebibliography}

\usepackage{graphicx}	% Including figure files
\usepackage{amsmath}	% Advanced maths commands
\usepackage{enumitem} %for numbers in enumerate

\title[Dust and ice during and after an outburst]{Collisional evolution of dust and water ice in protoplanetary discs during and after an accretion outburst}

\author[A. Houge, S. Krijt]{
Adrien Houge\thanks{E-mail: ah1006@exeter.ac.uk} and Sebastiaan Krijt
\\
Department of Physics and Astronomy, University of Exeter, Exeter, EX4 4QL, UK\\}

\date{Submitted: 19 December 2022, Accepted: 16 March 2023}

\pubyear{2023}

\begin{document}
\label{firstpage}
\pagerange{\pageref{firstpage}--\pageref{lastpage}}
\maketitle

% Abstract of the paper
\begin{abstract}

Most protoplanetary discs are thought to undergo violent and frequent accretion outbursts, during which the accretion rate and central luminosity are elevated for several decades. This temporarily increases the disc temperature, leading to the sublimation of ice species as snowlines move outwards. In this paper, we investigate how an FUor-type accretion outburst alters the growth and appearance of dust aggregates at different locations in protoplanetary discs. We develop a model based on the Monte Carlo approach to simulate locally the coagulation and fragmentation of icy dust particles and investigate different designs for their structure and response to sublimation. Our main finding is that the evolution of dust grains located between the quiescent and outburst water snowlines is driven by significant changes in composition and porosity. The time required for the dust population to recover from the outburst and return to a coagulation/fragmentation equilibrium depends on the complex interplay of coagulation physics and outburst properties, and can take up to $4500\mathrm{~yr}$ at $5\mathrm{~au}$. Pebble-sized particles, the building blocks of planetesimals, are either deprecated in water ice or completely destroyed, respectively resulting in drier planetesimals or halting their formation altogether. When accretion outbursts are frequent events, the dust can be far from collisional equilibrium for a significant fraction of time, offering opportunities to track past outbursts in discs at millimetre wavelengths. Our results highlight the importance of including accretion outbursts in models of dust coagulation and planet formation.

\end{abstract}


%%%%%%%%%%%%%%%%%%%%%%%%%%%%%%%%%%%%%%%%%%%%%%%%%%

%%%%%%%%%%%%%%%%% BODY OF PAPER %%%%%%%%%%%%%%%%%%

\section{Introduction}
\label{sec:introduction}

Planets form in discs of dust and gas around young stars. The first step of their formation occurs through the coagulation of the initial reservoir of sub-$\mu\mathrm{m}$-sized dust grains, aggregating into ${\approx} \mathrm{cm}$-sized pebbles through sticky collisions at low velocities \citep[][]{weidenschilling1993formation}. It is followed by the streaming instability which achieves the formation of km-sized planetesimals from dense clumps of pebbles. After that, gravity becomes the main driver of interactions to complete the formation of planets. The complication arises from the high sensitivity of the streaming instability towards grain size, as it requires sufficiently large grains to be triggered \citep[][]{bai2010dynamics, drkazkowska2014can, li2021thresholds}. Dust coagulation is thus a crucial step in the formation of planets, as its efficiency controls the occurrence of the following steps. Moreover, dust properties (e.g. composition, structure) are inherited by the planetesimals, so that understanding dust evolution allows to constrain the properties expected in larger objects \citep[][]{jansson2014formation}. Beyond the impact on planets, dust also has a key importance for the structure and evolution of protoplanetary discs, by dominating the absorption and scattering opacity in most regions \citep[][]{beckwith1990survey, beckwith1999dust, bouwman2000composition}, transporting volatiles both radially and vertically \citep[][]{cuzzi2004material, ciesla2006evolution, oberg2016excess, krijt2018transport}, and providing surface area to promote chemical reactions \citep[e.g.][]{kress2001role, ruaud2019three}.\\

However, modelling the growth and evolution of dust is a challenging task, as the efficiency of coagulation is related to the complex couplings of transport processes, disc conditions, and micro-physical properties of the dust grains. One notable example concerns the presence of water ice. In fact, it has been shown that ice-covered dust grains are characterised by a higher resistance towards fragmentation than bare rocks \citep[][]{supulver1997sticking, dominik1997physics, wada2013growth, gundlach2014stickiness}, allowing for the formation of larger pebbles in regions where water ice is stable, i.e. outside the water snowline \citep[][]{birnstiel2010gas, banzatti2015direct}. As a consequence, the efficiency of coagulation and the overall dust distribution vary dramatically across the snowline, which may offer a sweet spot for planetesimal formation \citep[e.g.][]{okuzumi2012rapid, drkazkowska2017planetesimal}. Dust evolution models furthermore predict a sharp change in the dust thermal emission at millimetre wavelengths as (1) grain sizes increase beyond the snowline and (2) reduced radial drift in the inner regions leads to a pile-up of small solids and an increase in the optical depth \cite{banzatti2015direct}. 

The position of the water snowline therefore plays a crucial role in dust evolution. However, protoplanetary discs can experience frequent accretion outbursts throughout their evolution, during which the accretion rate of the central protostar increases by ${\sim}2$ orders of magnitude and remains high for several decades \citep[][]{audard2014episodic}. Such events dramatically increase the temperature of the surrounding disc, pushing snowlines outward, and leading to the sublimation of ices on scales ${\geq} 10 \mathrm{~au}$. The impacts of outbursts on gas chemistry has been thoroughly studied, especially investigating whether some chemical tracers could be used to probe the occurrence of past outbursts in discs, to better constrain their causes and properties by enlarging our statistical sample of these events \citep[e.g.][]{molyarova2018chemical, wiebe2019luminosity}.

Outbursts and the ensuing sublimation of water ice in particular are expected to alter the dust size distribution. \citet{cieza2016imaging}, building on the work of \citet{banzatti2015direct}, used ALMA observations of dust emission (radial profiles of the optical depth and spectral index) to argue that the water snowline in the outbursting system of V883 Ori was located at 42 au. Depending on the duration of the outburst, however, it is not necessarily clear whether the situation is directly analogous to the models of \citet{banzatti2015direct} in which the snowline is static. First, the dust distribution needs time to respond to the new collisional equilibrium, and the dust pile-up is built up only after several radial drift timescales \citep{schoonenberg2017planetesimal}. Indeed, using a simplified monodisperse grain model, \citet{schoonenberg2017pebbles} showed that the features observed by \citet{cieza2016imaging} could be reproduced if ice-rich aggregates disintegrated following the outburst, and the remaining silicate grains were allowed to re-coagulate to sizes of approximately $300\mathrm{~\mu m}$.

The properties of the dust size distribution during and following the outburst thus depend sensitively on the mechanical response of the aggregates to losing their water, and on the details of the re-coagulation process. In this study, we investigate these processes in detail by performing local dust coagulation calculations in several specific locations of a disc undergoing a step change in temperature following an outburst. The aim is to quantify how the full dust size distribution responds to water ice leaving (when the outburst starts) and returning (soon after the end of the outburst) while at the same time undergoing collisional evolution. We also investigate the impact of different assumptions regarding the aggregate structure (e.g. compact vs. porous growth) and response to ice sublimation on the resulting dust size distribution and its (integrated) optical properties (e.g. mm spectral index). Similarly to what is done on gas tracers, we investigate whether the alteration of dust properties may offer an opportunity to track past outbursts in discs. 

This paper is organized as follows. In Sect. \ref{sec:discmodel}, we present the disc and outburst model used in this study. Dust properties, growth, and dynamics are then described in Sect. \ref{sec:dustmodel}. The collision model and Monte Carlo numerical approach to dust coagulation is presented in Sect. \ref{sec:simulationmethod}. Results of the coagulation simulations are presented in Sect. \ref{sec:results} along with their conversion into meaningful observational signatures in Sect.~\ref{sec:opacity_model}. The results are discussed in Sect. \ref{sec:discussion} followed by our conclusions in Sect. \ref{sec:conclusions}. Throughout this manuscript, dust of any size will be referred to as aggregates, solids, or particles. We will use ‘dust grains’ when specifically targeting small objects in the lower-end of the size distribution (i.e. $< 10 \mu \mathrm{m}$), 'pebbles' for the upper-end (i.e. $> 1 \mathrm{mm}$), and ‘population’ to describe the entire distribution.

\section{Disc model}
\label{sec:discmodel}

The gas surface density profile is based on a tapered power-law \citep[]{lynden1974evolution, hartmann1998accretion}
\begin{equation}
\Sigma_\mathrm{g} (r) = \Sigma_\mathrm{c} \bigg( \dfrac{r}{r_c} \bigg)^{-\gamma} \mathrm{exp} \bigg[ -\bigg( \dfrac{r}{r_c} \bigg)^{2-\gamma}\bigg],
\label{eq:surface_density}
\end{equation}

where $\Sigma_\mathrm{c}$ is the surface density normalization which is calculated from the total disc mass with
\begin{equation}
\Sigma_\mathrm{c} = \dfrac{M_\mathrm{disc}(2-\gamma)}{2\pi r_c^2}.
\label{eq:surface_density_integrated}
\end{equation}

The radial profile of the surface density is thus parameterised by three quantities, set to the following characteristic values: $r_\mathrm{c} = 100\mathrm{~au}$, $\gamma=1$, and $M_\mathrm{disc} = 0.01 ~M_*$ with $M_*=1~M_\odot$. The dust-to-gas ratio is set to $\delta_\mathrm{d2g} = 0.01$ and is assumed constant throughout the disc.

We assume the disc vertical structure to be in hydrostatic equilibrium, so that the vertical profile of the gas density is expressed as
\begin{equation}
\rho_{\mathrm{g}}(z)=\frac{\Sigma_{\mathrm{g}}}{\sqrt{2 \pi} h_{\mathrm{g}}} \exp \left\{-\frac{z^{2}}{2 h_{\mathrm{g}}^{2}}\right\},
\label{eq:gasdensity}
\end{equation}
where $h_\mathrm{g} = c_\mathrm{s}/\Omega$ is the gas pressure scale height, $c_\mathrm{s} = \sqrt{k_{\mathrm{B}} T / m_{\mathrm{g}}}$ is the sound-speed, $k_\mathrm{B}$ is the Boltzmann constant, $m_{\mathrm{g}} = 2.34$ amu is the mean molecular mass, and $\Omega=\sqrt{G M_{*} / r^{3}}$ is the Keplerian frequency.

The temperature of the disc midplane $T_\mathrm{m}(r)$, where our coagulation simulations take place, is connected to the amount of energy absorbed by the disc atmosphere from the central luminosity source (star and accretion region) and re-emitted downward. Neglecting viscous heating for simplicity, it is expressed as \citep[][]{chiang1997spectral}
\begin{equation}
\label{eq:midplane_temperature}
T_\mathrm{m}^4(r) = \dfrac{\phi(r)}{8 \pi\sigma_\mathrm{SB} r^2} (L_\mathrm{*}+L_\mathrm{acc}),
\end{equation}
where
\begin{equation}
\label{eq:flaringangle}
\phi(r) \simeq \frac{0.4 R_{*}}{r} ~+~r \frac{ \mathrm{~d}\left(h_{\mathrm{p}} / r\right)}{\mathrm{~d}r},
\end{equation}
represents the disc opening angle, related to the scale height of the visible photosphere 
\begin{equation}
\label{eq:photospherescaleheight}
h_\mathrm{p} = h_{0} \left(\frac{r}{r_0}\right)^{\Psi}.
\end{equation}
We set the stellar radius to $R_{*}=2.5 ~R_\odot$, the disc flaring index to $\Psi=1.26$, and the scale height to $h_{0}=34.2\mathrm{~au}$ at $r_{0}=100 \mathrm{~au}$ \citep[][]{benisty2022optical, lagage2006anatomy}. 

For our young solar-mass star, we set the stellar luminosity to $L_\mathrm{*} = 0.9~L_\odot$ and the contribution of the quiescent accretion region to $L_\mathrm{acc} = 0.3~L_{\odot}$ \citep[][]{molyarova2018chemical}. The former is assumed constant while the later will vary during episodic outbursts. With our stellar parameters, the value for the accretion luminosity is comparable with an accretion rate of $\dot{M} = (2/3) (L_\mathrm{acc} R_*/ G M_*) \approx 10^{-8}~M_{\odot}/\mathrm{yr}$ , which is consistent with observed values \citep[]{audard2014episodic}. Note that we assume the gas and dust temperature to be equal, which is a valid assumption in the dense midplane region.

\subsection{Outburst event}
\begin{figure}
	\includegraphics[width=\columnwidth]{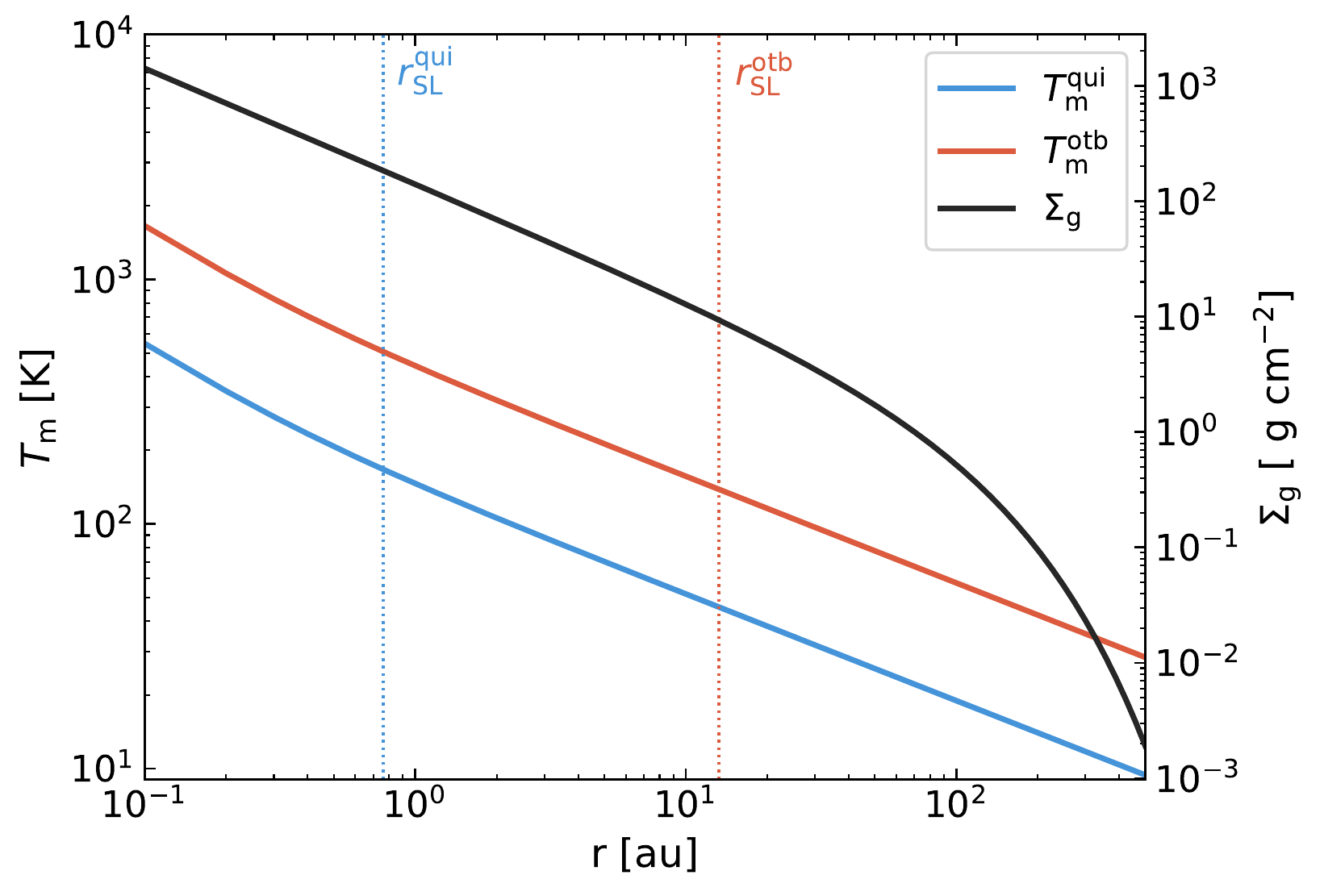}
    \caption{Midplane temperature and gas surface density radial profile. The vertical dotted lines denote the position of the quiescent and excited water snowlines.}
    \label{fig:fig1}
\end{figure}

We introduce an FUor-type accretion outburst in our quiescent system at $ t_\mathrm{otb}^\mathrm{start} = 10^4 \mathrm{~yr}$, in agreement with current knowledge of such outburst rates \citep[e.g.][]{scholz2013systematic}. For our purposes, we mimic a single outburst by raising the accretion luminosity to $L_\mathrm{acc} = 100~L_\odot$ for a duration $\tau_\mathrm{otb} = 100\mathrm{~yr}$. The temperature of the disc midplane increases according to Eq.~\ref{eq:midplane_temperature} (see Fig.~\ref{fig:fig1}), and we assume it adapts instantaneously as the heating timescale of the disc is short as compared to the outburst duration \citep[][]{johnstone2013continuum, vorobyov2014influence}. We further assume the gaseous environment to instantaneously find a new hydrostatic equilibrium in the vertical direction when the temperature is modified, which leads to slightly lower midplane gas density during the outburst. This assumption is valid given that the thermal timescale of the gas in our disc model is inferior to the dynamical timescale and outburst duration. Using Equation (7) from \citet[][]{ueda2021thermal}, we find $\sim 5\mathrm{~yr}$ at $5\mathrm{~au}$. The solid density follows the same behaviour to maintain the dust-to-gas ratio to its fixed value. For simplicity, we do not consider potential increases of the disc opening angle due to flaring effects.

\subsection{Water content}
\label{sec:watercontent}

By raising the temperature, the outburst drives the sublimation of ices over extended regions of the disc, pushing outward the snowlines of various molecular species. In particular, water is an important compound in terms of abundance \citep[][]{lodders2003solar}, and its presence or absence in ice phase has a dramatic impact on dust growth as it influences the stickiness of aggregates \citep[][]{supulver1997sticking, dominik1997physics, wada2013growth, gundlach2014stickiness}, allowing for the formation of larger pebbles in regions where water ice is stable \citep[][]{birnstiel2010gas}.

Upon ice sublimation, it is still unclear how the structure of aggregates is impacted, as laboratory experiments found it could lead to both a complete disruption \citep[][]{aumatell2011breaking} or survival \citep[][]{spadaccia2022fate}. To fit the range of possibilities, we adopt two designs for the response to sublimation when the outburst starts: the "resilient" where aggregates survive and are just impacted by the loss of their ice mass, and "many-seeds" where all aggregates disrupt to monomer size as we consider water ice to 'glue' refractory grains together. The many-seeds model was also used by \citet[]{schoonenberg2017planetesimal} in the context of pebbles drifting inward and crossing the snowline. 

The water snowline is located where the sublimation and condensation rates of $\mathrm{H_2O}$ molecules have similar absolute values. They are given respectively by \citep[e.g][]{supulver2000formation}
\begin{equation}
    F_\mathrm{sub} = -\sqrt{\dfrac{m_\mathrm{H_2O}}{2\pi k_\mathrm{B}T}} P_\mathrm{sat},
\end{equation}
\begin{equation}
    F_\mathrm{con} = \sqrt{\dfrac{m_\mathrm{H_2O}}{2\pi k_\mathrm{B}T}} P_\mathrm{H_2O},
    \label{eq:condensation_flux}
\end{equation}
with $P_\mathrm{H_2O}$ the water vapour pressure expressed with the ideal gas law as
\begin{equation}
    P_\mathrm{H_2O} = \dfrac{k_\mathrm{B}T}{m_\mathrm{H_2O}} \rho_\mathrm{H_2O},
\end{equation}
and $P_\mathrm{sat}$ the saturated vapour pressure for water on a flat surface \citep[]{supulver2000formation} given by 
\begin{equation}
\begin{split}
P_\mathrm{sat} & = \dfrac{k_\mathrm{B}T}{m_\mathrm{H_2O}} \rho_\mathrm{sat}, \\
 & = 1.013 \times 10^6 ~\exp \Bigg\{ 15.6 - \dfrac{5940 K}{T} \Bigg\} ~\mathrm{dyn/cm^{2}},     
\end{split}
\end{equation} 
where $m_\mathrm{H_2O}$ is the mass of an $\mathrm{H_2O}$ molecule, and $\rho_\mathrm{H_2O} = \delta_{\mathrm{w2g}} \rho_\mathrm{g}$ assuming the water abundance to $\delta_{\mathrm{w2g}}=0.01$. With this assumption, the dust-to-ice ratio equals unity in the outer disc.

\begin{figure}
	\includegraphics[width=\columnwidth]{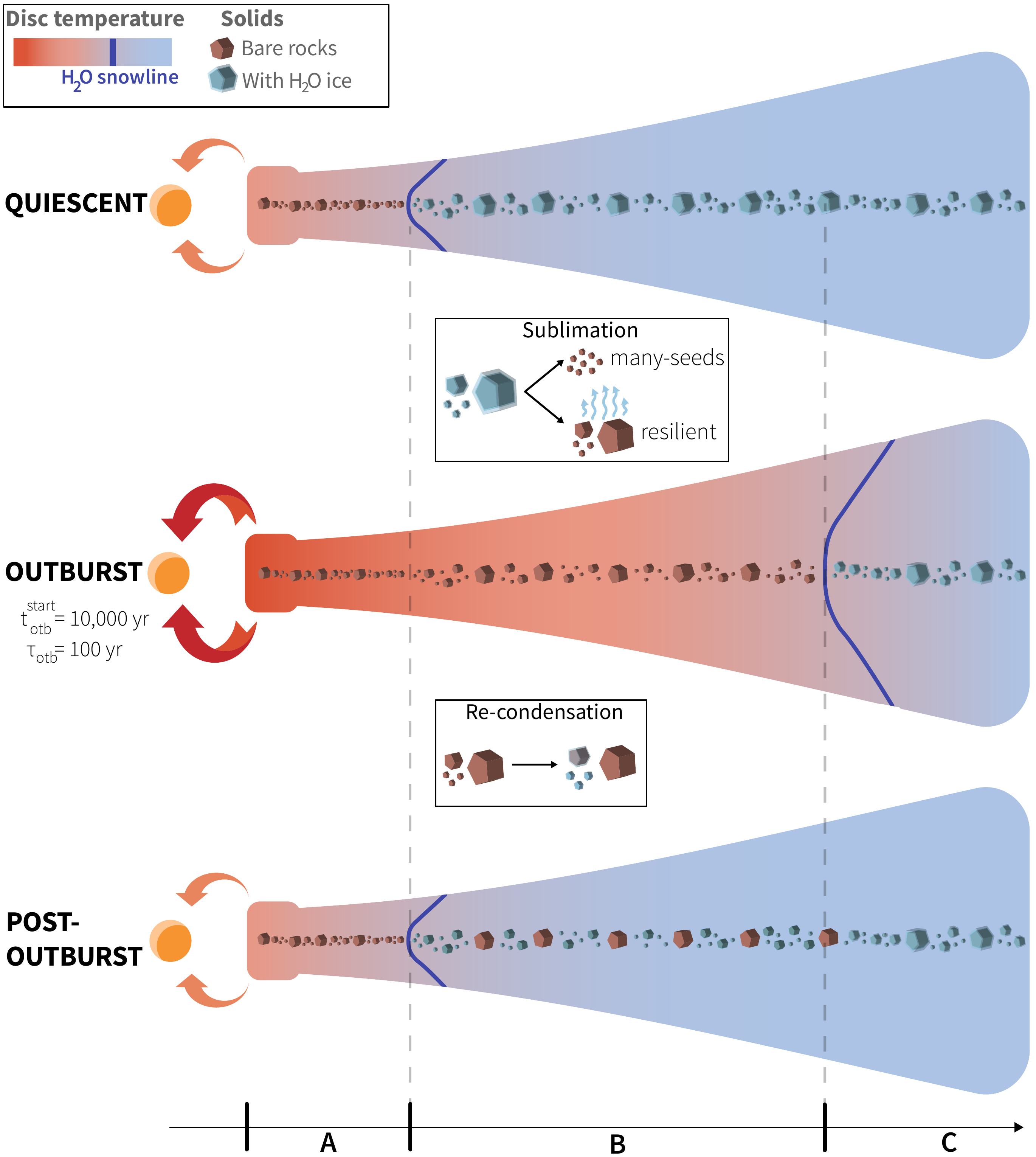}
    \caption{Cartoon representing the thermal structure and dust grains of a protoplanetary disc undergoing an accretion outburst. Three zones are specified. Zone A: the temperature is too high for water to be stable in ice phase, thus remaining in gas phase at all time. Zone B: water is initially deposited onto refractory cores, enhancing their stickiness and favouring growth. It sublimates during the outburst, and re-condensates after the event proportionally to the total dust surface area. Depending on the model, dust grains survive or not the sublimation of water. Zone C: even during the outburst, the temperature is low enough for water to remain in ice phase.}
    \label{fig:cartoon}
\end{figure}

In our disc model, the quiescent water snowline is located at $r_\mathrm{SL}^\mathrm{qui} = 0.8 ~\mathrm{AU}$ from the central protostar, corresponding to a temperature of $167~\mathrm{K}$ (see Fig.~\ref{fig:fig1}). It is pushed at $r_\mathrm{SL}^\mathrm{otb} = 13 ~\mathrm{AU}$, during the accretion outburst, corresponding to a lower sublimation temperature of $122 ~\mathrm{K}$ as the gas density also decreases with the distance to the star (equation~\ref{eq:gasdensity}). The snowline during the outburst will be further referred to as the excited snowline. As illustrated in Fig.~\ref{fig:cartoon}, we can now divide the disc into three zones: A) inside the quiescent snowline, water is in a vapour phase at all time; B) in between the quiescent and excited snowline, the phase of water molecules will vary with the outburst; and C) outside the excited snowline, water always remains in an ice state. We will perform local dust coagulation simulations in the midplane within zone A and B\footnote{We will come back to zone C in Sect. \ref{sec:results_compact_A}.}, respectively at $0.5$ and $5\mathrm{~au}$, hereafter referred to as location A and B. The local disc conditions can be found in Table \ref{tab:locations_values}. Our approach being local and in the midplane, we do not include the potential transport of material due to vertical settling and radial drift (see Sect.~\ref{sec:radialdriftlimitation}). 

\begin{table}
	\centering
	\caption{Local disc parameters in location A and B during the quiescent and outburst phases.}
	\label{tab:locations_values}
	\begin{tabular}{lll} % four columns, alignment for each
		\hline
		Local parameter & \textbf{A} & \textbf{B}  \\
		\hline
		Heliocentric distance $r$ (au) & 0.5 & 5 \\
		Gas surface density $\Sigma_g$ ($\mathrm{g~cm^{-2}}$) & $281.48$ & $26.9$ \\
		\hline
		\textbf{Quiescent phase} & & \\
		$\rho_\mathrm{g}$ ($\mathrm{g~cm^{-3}}$) & $7.4 \times 10^{-10}$ & $3.8 \times 10^{-12}$ \\
		$T_\mathrm{m}$ (K) & 207.06 & 70.13 \\
		$\eta$ & 0.0006 & 0.0017 \\	
		\hline
		\textbf{Outburst phase} & & \\
		$\rho_\mathrm{g}$ ($\mathrm{g~cm^{-3}}$) & $4.2 \times 10^{-10}$ & $2.2 \times 10^{-12}$ \\

		$T_\mathrm{m}$ (K) & 627.01 & 212.35 \\
		$\eta$ & 0.0017 & 0.0058 \\

		\hline
	\end{tabular}
\end{table}

\section{Dust models}
\label{sec:dustmodel}

\subsection{Monomers}
\label{sec:monomers}
In protoplanetary discs, solids initially consist of sub-$\mu$m-sized dust grains, referred to as monomers, whose motion is well coupled to the surrounding gas. In this work, we assume initially two distinct monomer populations based on their position in the quiescent disc. Inside the quiescent water snowline, monomers are chosen to be identical $a_0 = 0.1 \mu\mathrm{m}$ compact spheres made of a rocky mix of silicates, troilite, and refractory organics (see Table \ref{tab:dustcomposition}). The bulk density of the mixture is $\rho_\mathrm{s, < SL} = 2.11  ~\mathrm{g~cm^{-3}}$. Outside the quiescent water snowline, $\mathrm{H_2O}$ molecules are accreted onto dust grains. In that case, we assume water is homogeneously mixed with the rock mix such that the water mass fraction $f_\mathrm{w} = m_\mathrm{w}/(m_\mathrm{w}+m_\mathrm{r}) = 0.5$ \citep[]{lodders2003solar}. The bulk density is then $\rho_\mathrm{s, > SL} = 1.28  ~\mathrm{g~cm^{-3}}$.

\begin{table}
	\centering
	\caption{Monomer material density and relative mass abundance $f_i$ for the two populations \citep[]{birnstiel2018disk, lodders2003solar}}
	\label{tab:dustcomposition}
	\begin{tabular}{lccc} % four columns, alignment for each
		\hline
		Material & Density $[ ~\mathrm{g~cm^{-3}}]$ & $f_{i,\mathrm{< SL}}$ & $f_{i, \mathrm{> SL}}$ \\
		\hline
		Silicates & 3.30 & 0.411 & 0.206 \\
		Troilite & 4.83 & 0.093 & 0.046 \\
		Refractory organics & 1.50 & 0.496 & 0.248 \\
		Water ice & 0.92 & 0 & 0.5 \\
		\hline
	\end{tabular}
\end{table}

\subsection{Dust dynamics}
\label{sec:dustdynamics}

Due to the interaction with the surrounding gaseous environment, dust grains acquire non-zero relative velocities, leading to their coagulation into larger and larger aggregates. The aerodynamic behaviour of such embedded solids is quantified with the Stokes number $ \mathrm{St} = \Omega t_\mathrm{s}$, where $t_\mathrm{s}$ is the stopping time. For the Epstein and Stokes drag regime, it is expressed as \citep[][]{okuzumi2012rapid}
\begin{equation}
\label{eq:stokesnumber}
        t_\mathrm{s}=\left\{\begin{array}{ll}
t_\mathrm{s}^{(\mathrm{Ep})} \equiv \dfrac{3 m}{4 \rho_\mathrm{g} v_{\mathrm{th}} A}, & a<\frac{9}{4} \lambda_{\mathrm{mfp}}, \\\\
t_\mathrm{s}^{(\mathrm{St})} \equiv \dfrac{4 a}{9 \lambda_{\mathrm{mfp}}} t_\mathrm{s}^{(\mathrm{Ep})}, & a \geq \frac{9}{4} \lambda_{\mathrm{mfp}},
\end{array}\right.
\end{equation}
where $v_\mathrm{th} = \sqrt{8/\pi}c_\mathrm{s}$ is the thermal velocity, $\lambda_\mathrm{mfp} = m_\mathrm{g}/(\sigma_\mathrm{mol} \rho_\mathrm{g})$ is the mean free path of gas particles, $\sigma_\mathrm{mol} = 2 \times 10^{-15} ~\mathrm{cm^{3}}$ is the collision cross section of gas molecules, and $A$ is the projected surface area of the aggregate.

We consider in our simulations the typical sources of relative velocities, namely the Brownian motion, the turbulence \citep[based on equation~16 of][]{ormel2007closed}, and the radial and azimuthal drifts \citep[see Sect. 3.1 of][]{birnstiel2016dust}. The turbulent motion is parametrized using the $\alpha$-turbulence model of \citet[][]{shakura1973black}, where we assume the turbulence strength to a constant value $\alpha=10^{-3}$ \citep[][]{rosotti2023empirical}.
As we restrict our study to the midplane, the velocity arising from vertical settling is zero. Drifting motions depend on $\eta$, the dimensionless radial pressure gradient, expressed as
\begin{equation}
2 \eta \equiv-\left(\frac{c_\mathrm{s}}{v_\mathrm{K}}\right)^{2} \frac{\partial \ln \left(\rho_\mathrm{g} c_\mathrm{s}^{2}\right)}{\partial \ln r}.
\end{equation}
Its local value during the quiescent and outburst phase is given in Table \ref{tab:locations_values}. As previously stated, we neglect the potential transport of material due to radial drift, but we do consider its impact as a relative velocity source. 

\subsection{Aggregation}
\label{sec:aggregation}

The local dust coagulation process depends sensitively on the structure of the growing grains for a variety of reasons. For example, substantial porosity impacts the aggregate mass-size relation \citep[][]{blum2000growth}, affecting its collisional cross section, its aerodynamical behavior (Eq.~\ref{eq:stokesnumber}), and its ability to dissipate energy during collisions \citep[][]{blum2008growth}. Furthermore, the appearance of the aggregate (i.e. its opacity at different wavelengths) is a sensitive function of porosity \citep[][]{kataoka2014opacity}. Models of dust coagulation in planet-forming environments are somewhat split, with traditional approaches assuming compact, spherical particles at all times, while models that include porosity evolution have reported internal grain densities as low as $\mathrm{10^{-5} ~g~cm^{-3}}$ \citep[][]{okuzumi2012rapid}. To explore the possible range of outcomes we will contrast two different cases: compact coagulation (Sect. \ref{sec:compactaggregation}) and porous growth (Sect. \ref{sec:porousaggregation}).

\subsubsection{Compact growth}
\label{sec:compactaggregation}

The compact model assumes aggregates to keep a compact homogeneous spherical shape throughout their growth. In that case, an aggregate's size $a$ and mass $m$ are connected through $m = (4/3)\pi a^3 \rho_\mathrm{s}$, where the bulk density remains equal to their constituting material. As demonstrated by laboratory experiments \citep[]{blum2008growth, guttler2010outcome}, there exists a multitude of collision outcomes depending on the colliders composition, relative velocity, and mass ratio. In this work, we will only consider perfect sticking (leading to growth), and fragmentation (leading to mass loss), as there is still a large parameter space to be explored concerning other outcomes (e.g. bouncing, erosion). Fragmentation occurs if the relative velocity is above the fragmentation limit $v_\mathrm{f}$ and if the mass ratio $R_\mathrm{m}$ of the colliders is superior to $R_\mathrm{m, crit} = 0.01$ \citep[][]{guttler2010outcome, seizinger2013erosion}. We set the fragmentation velocity to $1\mathrm{~m~s^{-1}}$ for bare rock material and $10\mathrm{~m~s^{-1}}$ for pure water ice, in agreement with laboratory experiments finding enhanced stickiness for water-rich solids \citep[e.g][]{supulver1997sticking, gundlach2014stickiness}. As in our case aggregates are rather homogeneously mixed in ice and rock, we express their fragmentation velocity as a linear interpolation between the pure rock and ice cases \citep[][]{lorek2016comet}
\begin{equation}
    \label{eq:vf_interpolation}
    v_\mathrm{f} = f_\mathrm{w} v_\mathrm{f}^\mathrm{H_2O} + (1-f_\mathrm{w})v_\mathrm{f}^\mathrm{rock}.
\end{equation}

We note that these values are now under debate in the light of recent experiments on the resistance of water ice grains at low temperatures \citep[][]{gundlach2018tensile, musiolik2019contacts}.

\subsubsection{Porous growth}
\label{sec:porousaggregation}

In the porous model, we include the evolution of the dust aggregate's structure (i.e. porosity), as in reality aggregates can develop a significant fractal shape which alters the mass-size relation following $m \propto a^{D_\mathrm{f}}$, $D_\mathrm{f}$ being the fractal dimension, and leads to a much smaller internal density \citep[e.g.][]{donn1990formation, blum2000growth, weidling2009physics}. On the microscopic level, aggregates are considered to be build up of monomers whose properties and bonds dictate the mechanical behaviour of the aggregate as a whole. Because contact between microscopic spheres only involves a small surface layer of relative thickness $\delta \approx 10^{-2}$ \citep[e.g.][]{chokshi1993dust, krijt2013energy}, only a small fraction of water ice is needed to alter the surface properties from bare rock to those of pure water. Similarly to \citet[]{krijt2016tracing}, we define that mass fraction threshold to be $f^* = m_\mathrm{w}/m_\mathrm{r} = 0.1$. 

We use the porosity model from \citet{okuzumi2012rapid} to calculate the new aggregate volume after every sticky collision, considering the creation of new voids and the potential collisional compression. The efficiency of collisional compression is controlled by how the impact energy $E_\mathrm{imp}$ compares to the rolling energy $E_\mathrm{roll}$, which quantifies the ability of monomers in contact to roll over each other \citep[]{dominik1997physics}. Using $f^*$ to characterise the surface properties of monomers, the rolling energy is given by\footnote{We choose to use the rolling energy of $\mathrm{SiO_2}$ to represent our rocky composition.} \citep[]{heim1999adhesion, gundlach2011micrometer, krijt2014rolling} 
\begin{equation}
E_{\mathrm{roll}} = \left\{\begin{array}{ll}
E_{\mathrm{roll}}^{\mathrm{H}_{2} \mathrm{O}} = 1.4 \times 10^{-7} \operatorname{erg}\left(a_{0} / \mu \mathrm{m}\right)^{5 / 3}, & f^* > 0.1, \\\\
E_{\mathrm{roll}}^{\mathrm{rock}} = 2.3 \times 10^{-8} \operatorname{erg}\left(a_{0} / \mu \mathrm{m}\right)^{5 / 3}, & f^* < 0.1,
\end{array}\right.
\end{equation}
At small sizes, when the relative velocity is governed by Brownian motion and $E_\mathrm{imp} < E_\mathrm{roll}$, there is no dissipation of energy through restructuration which results in gentle hit-and-stick collisions and a low fractal dimension $D_f \simeq 2$ \citep[e.g.][]{kempf1999n}. With increasing mass and impact energy, collisional compression occurs which increases the average density of dust aggregates. When the aggregates are so large their motion decoupled from the gas flow, the efficiency of collisional compression stalls, allowing the formation of highly porous aggregates with $\rho_\mathrm{int} \approx 10^{-5} \mathrm{~g~cm^{-3}}$ \citep[][]{okuzumi2012rapid}. However, \citet[]{kataoka2013fluffy} argued that static compression by gas ram pressure and self-gravity would prevent the formation of such massive and highly porous solids. We thus added their prescription to our coagulation model (see Sect.~\ref{sec:selfcompression}).

Similarly to the compact case, colliders with a mass ratio $R_\mathrm{m} \geq 0.01$ fragment if their relative velocity is above the fragmentation limit $v_\mathrm{f}$. For porous aggregates, numerical simulations of individual collisions show $v_\mathrm{f}$ depend on the monomer properties and can be as high as $80 \mathrm{~m~s^{-1}}$ for $0.1$ $\mu$m pure water ice \citep[][]{wada2013growth}. For such high $v_\mathrm{f}$, the fragmentation threshold is never reached and direct growth to planetesimals may be possible in some regions of discs \citep[][]{okuzumi2012rapid}. Here we use a slightly more conservative version of the results from \citet[]{wada2013growth}:
\begin{equation}
v_\mathrm{f} \simeq \left\{\begin{array}{ll}
30\left(\frac{a_{0}}{0.1 \mu \mathrm{m}}\right)^{-5 / 6} \mathrm{~m} \mathrm{~s}^{-1}, & f^* > 0.1, \\\\
3\left(\frac{a_{0}}{0.1 \mu \mathrm{m}}\right)^{-5 / 6} \mathrm{~m} \mathrm{~s}^{-1}, & f^* < 0.1.
\end{array}\right.
\label{eq:frag_limit_porous}
\end{equation}
Although higher than the $v_\mathrm{f}$ for the compact case, these values still result in fragmentation-limited growth outside the snowline.\\

Another aspect to outline concerning porous growth is that the relation for the surface area $A = \pi a^2$ can break down for fractal aggregates \citep[]{okuzumi2009numerical, tazaki2021analytic}, we rather adopt a corrected definition as formulated by Equation~(47) of \citet[]{okuzumi2009numerical}.

\section{Numerical method}
\label{sec:simulationmethod}

\subsection{Superparticle approach}
\label{sec:superparticleapproach}

We are simulating the coagulation and fragmentation of a population of dust grains using the Monte Carlo superparticle approach from \citet[]{zsom2008representative}. It follows the evolution of $n=10^4$ superparticles, each one representing a large swarm of physical particles with identical properties. The total rock mass of each swarm, $M_\mathrm{swm}$, is fixed, so that if the rocky content of a superparticle $i$ changes, the number of physical particles it represents is modified following $N_i = M_\mathrm{swm} / m_{\mathrm{r}}$. The particles are distributed evenly in a fixed volume $V$. The water content is treated apart from the swarm consideration, to ensure the conservation of the number of particles when only the water mass changes, i.e. upon sublimation and condensation. It leads to small statistical fluctuations in the total water mass, which we discuss in Sect.~\ref{sec:conserving_water}. The particle properties we follow are: the mass of rock ($m_\mathrm{r}$) and water ice ($m_\mathrm{w}$), the size $a$, and the porosity in the porous case through the internal density $\rho_\mathrm{int}$.

The coagulation code works following four key steps. First, we calculate the collision rates between every pair of particles. The collision rate of a superparticle $i$ with a physical particle represented by the superparticle $j$ is given by $C_{ij} = N_j \Delta v_{ij} \sigma_{ij} / V$, where $\sigma_{ij} = \pi (a_i + a_j)^2$ is the collisional cross-section, and $\Delta v_{ij}$ is the relative velocity calculated from the motion processes mentioned in Sect. \ref{sec:dustdynamics}. 

Then, two random numbers determine which superparticle $i$ will collide with which physical particle of the swarm $j$, such that pairs with large collision rates are more likely to be drawn. In addition, using the total collision rate 
\begin{equation}
 C_\mathrm{tot} = \sum\limits_{i,j} C_{ij},
\end{equation}
and a random number $\mathcal{R}$ drawn from a uniform distribution between $0$ and $1$, we determine the time-step to that next collision as
\begin{equation}
\delta t_{\mathrm{col}}=-\frac{\ln \mathcal{R}}{C_{\mathrm{tot}}}.
\end{equation}
After that, the collision outcome (sticking or fragmentation) is determined based on the colliders mass ratio $R_\mathrm{m}$ and the relative velocity $\Delta v_{ij}$ as compared to the the fragmentation threshold $v_\mathrm{f}$ (see Sect.~\ref{sec:aggregation}). We employ the collision model from \citet{birnstiel2011dust}
to account for the intermediate regime between sticking and fragmentation, using a width $\delta v_\mathrm{f}=v_\mathrm{f}/5$, as experimental results did not reveal a sharp transition \citep[]{blum1993experimental}. Finally, we update the properties of the superparticle $i$ in agreement with the collision outcome, while those of the physical particle $j$ it collides with are left unchanged \citep[][]{zsom2008representative}.  

These four steps represent an individual collision cycle, during which the global time of the simulation is incremented by $\delta t_\mathrm{col}$. The coagulation code repeats the cycle until it reaches $t_\mathrm{end}=10^5 \mathrm{~yr}$, time threshold fixed by the user. Because the particles properties are constantly monitored, we can analyse their evolution and distribution amongst the population at any chosen time.

\subsubsection{Grouping method}

The size distribution can be broad, especially when the fragmentation barrier is reached and collisions create a second generation of dust grains. As a consequence, collisions may involve a particle $j$ considerably less massive than its pair $i$, hence outcoming on minuscule changes for $i$. For numerical optimization purposes, we rather form a group of $j$-particles of mass $f_\mathrm{c} m_i$, and modify the corresponding collision rate to $\widetilde{C}_{ij} = C_{ij} f_\mathrm{c}^{-1} m_j / m_i$, where we set the grouping limit to $f_\mathrm{c}=0.01$ \citep[e.g.][]{zsom2008representative, krijt2016dust}. 
Doing so, the superparticle $i$ has less probability to encounter the group of $j$-particles, but when it does, it collides with all the particles of that group at once.

\subsection{Collisional evolution}
\label{sec:collisionalevolution}
When a collision takes place, the superparticle properties are modified depending on the selected collision outcome (sticking or fragmentation). In this section, we detail how each collision outcome modifies particles properties in both aggregation models. We refer the reader to Fig.~\ref{fig:cartoon_dustmodel} for a cartoon summarising our dust model.

\subsubsection{Sticking}

When colliding partners stick, the mass of the superparticle is updated to the sum of the colliders mass $m \mapsfrom m_i+m_j$. If they have different composition, the water mass fraction $f_\mathrm{w}$ and material density $\rho_\mathrm{s}$ are updated accordingly. In the compact case, we then determine the new size directly from $a = (3m/4\pi \rho_\mathrm{s})^{1/3}$. In the porous model, the size is replaced by the notion of characteristic size, $a_\mathrm{c}$, which is defined using the gyration radius of the aggregate \citep{mukai1992}. This radius can be used to define an aggregate's volume using $V=(4/3)\pi a_\mathrm{c}^3$. The new volume is computed after each collision using Eq. (15) from \citet[]{okuzumi2012rapid}, which accounts for the creation of new voids as well as possible collisional compression. The new mass and volume yield an internal density $\rho_\mathrm{int}=m/V$ from which we assess the porosity of aggregates. Note that if the colliding aggregates have different surface properties, the rolling energy is taken as the mass-weighted average. 

\subsubsection{Catastrophic fragmentation}
\label{sec:fragmentation}

While each aggregate's fragmentation velocity is set by its composition, the effective fragmentation limit of a colliding pair is obtained from their mass-weighted average. At higher velocities, and provided the collider mass ratio $R_\mathrm{m} \geq 0.01$ (see Sect.~\ref{sec:aggregation}), catastrophic fragmentation occurs and the mass of both colliders is redistributed over fragments. The fragment distribution follows a power-law function
\begin{equation}
n_{\mathrm{f}}(m)= \begin{cases}C_{\mathrm{f}} m^{-\xi} & \text { for } m_0 \leqslant m \leqslant m_{\mathrm{f}, \max }, \\ 0 & \text { else, }\end{cases}
\end{equation}
where $m_0$ is the monomer mass, $m_{\mathrm{f}, \max }$ is the mass of the largest fragment being the larger collider, and $C_\mathrm{f}$ is a constant equals to the sum of both collider masses. The power-law of the distribution is set to $\xi = \frac{11}{6}$ similarly to \citet[]{birnstiel2011dust}, such that the surface area is dominated by the smaller fragments, while the larger dominate the total mass. A random number is used to draw a single fragment mass amongst the distribution. 

In the compact case, we then yield the selected fragment's size using the sphere equation similarly to the sticking outcome prescription. For porous aggregates, we assume that the fragment internal density follows the historical evolution of its predecessor, i.e. it remains constant unless the resulting volume is larger than what would be expected in the hit-and-stick regime, $V_\mathrm{h\&s}$, in which case the volume is set to this value and the internal density is adjusted.

\subsection{Non-collisional evolution}
In this section, we detail how we treat changes in aggregates properties due to processes unrelated to their collisional evolution. Again, we refer the reader to Fig.~\ref{fig:cartoon_dustmodel} for a cartoon summarising our dust model.

\subsubsection{Gas and self-gravity compression}
\label{sec:selfcompression}
As mentioned in Sect. \ref{sec:porousaggregation}, the internal density of porous aggregates can be increased due to static compression by gas ram pressure and self-gravity. We implement this effect in our coagulation simulation following \citet[][]{kataoka2013fluffy}, by calculating the compressive strength of aggregates whenever their properties are modified due to collisions or the outburst (changing disc conditions, sublimation, condensation), and compressing them if they cannot withstand the aforementioned external pressures. 

\subsubsection{Sublimation and condensation}
\label{sec:sublimation&condensation}

We now detail how we treat the modification of aggregates properties upon water sublimation (for both resilient and many-seeds models), and re-condensation. We begin with the case of compact aggregates. Upon sublimation, for the many-seeds model, aggregates instantaneously disintegrate to rocky monomers (see Table \ref{tab:dustcomposition}). For the resilient model, they lose half of their mass, as initially $f_\mathrm{w}=0.5$, and we assume the rocky left-overs to remain compact spheres. The variation in material density and size are determined accordingly to the mass loss. Upon re-condensation at the end of the outburst, water is distributed between aggregates of different sizes proportionally to their surface area (see equation (\ref{eq:condensation_flux})), updating the properties of all particles under the assumption they remain spherical and homogeneous ice/rock mixtures. We assume the total condensing water mass to be equal to what sublimated at $t_\mathrm{otb}^\mathrm{start}$, neglecting potential losses through e.g. diffusion/advection or  gas-phase chemical reactions. Unlike sublimation, the freeze-out of molecules on grain surfaces can take a considerable amount of time, especially in the outer disc \citep[e.g. $10^3 -10^4 \mathrm{~yr}$ for CO molecules, see][]{vorobyov2013effect}. It is quantified by the freeze-out timescale $\tau_\mathrm{f}$, which depends on the gas density and size distribution. We evaluate $\tau_\mathrm{f}$ of $\mathrm{H_2O}$ molecules on compact and porous aggregates as
\begin{equation}
    \label{eq:freezeout}
    \tau_\mathrm{f} = \left( \frac{v_{\mathrm{th}}}{V} \sum_{i} \dfrac{M_i}{m_i} A_{i}\right)^{-1},
\end{equation}
where $M_i$ is the total mass of particles at a given mass $m_i$, and $A_i$ is their surface area. This expression falls back to the timescale given by Eq. (26) in \citet[]{krijt2016tracing} for the compact case. In our simulations, the freeze-out timescale is typically below $10\mathrm{~yr}$, so we assume re-condensation to be instantaneous at $t_\mathrm{otb}^\mathrm{end}$. We note that if condensation onto small grains is inefficient, for example because of grain curvature \citep[][]{sirono2011sintering}, or if the slow cooling rate leads to preferential condensation on a favourable grain size \citep[][]{hubbard2016fu}, ices will not accrete freely on the entire distribution and may boost ice formation in a specific size range. Such effects are not included in this work.\\

\begin{figure*}
	\includegraphics[width=\textwidth]{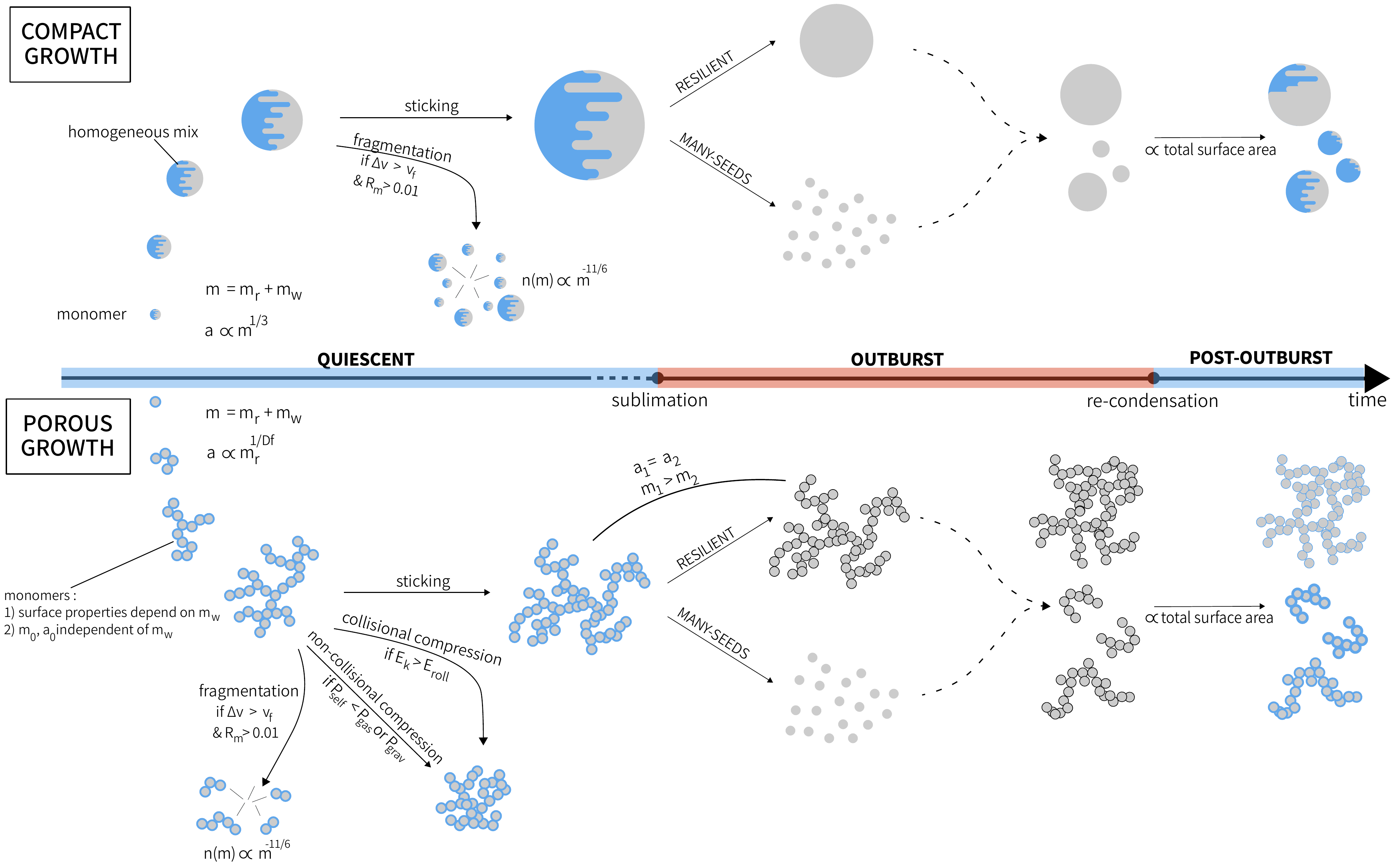}
    \caption{Cartoon summarising how compact and porous dust aggregates evolve in our simulations, including: coagulation, fragmentation, compression, sublimation, and condensation. In the compact model, monomers grow into sphere homogeneously mixed in ice and rock. Aggregates remain spherical throughout their evolution, even after sublimation and condensation. In the porous case, a few assumptions are made to avoid dealing with multiple monomer properties within a single aggregate. Water ice influences the aggregate surface properties, internal density, mass and inertia, but has no impact on its size and volume (except when disintegrating in the many-seeds model).}
    \label{fig:cartoon_dustmodel}
\end{figure*}

In the porous case, it is more complex to capture the structural impact of sublimation and condensation as it may lead to aggregates constituted of monomers with heterogeneous properties, for which the mechanical and collisions properties are not well known. We therefore make the following assumption: First, we only consider the impact of water on the monomer surface properties (i.e. their stickiness and rolling energy), and neglect the influence of water ice on the monomers' mass and size. Then, we add the mass of condensed water when calculating macroscopic aggregate properties such as total mass, mean density, Stokes number, etc., while using the rocky component to define the total volume and size (see bottom row of Fig.~\ref{fig:cartoon_dustmodel}).

\section{Results}
\label{sec:results}

In this section we describe the results for the different locations and dust models. To facilitate discussions, we will refer to each model following the notation introduced in Table \ref{tab:simulation_notation}.

\begin{table}
	\centering
	\caption{Model notations depending on the disc location, aggregation, and response to sublimation.}
	\label{tab:simulation_notation}
	\begin{tabular}{lccc} % four columns, alignment for each
		\hline
		Model & Location & Aggregation & Sublimation \\
		\hline
		\texttt{A-comp} & A & Compact & $\times$  \\
		\texttt{A-por} & A & Porous & $\times$ \\
		\texttt{B-comp-resi} & B & Compact & Resilient \\
		\texttt{B-comp-m.s} & B & Compact & Many-seeds \\
		\texttt{B-por-resi} & B & Porous & Resilient \\
		\texttt{B-por-m.s} & B & Porous & Many-seeds \\
		\hline
	\end{tabular}
\end{table}

\subsection{Compact growth}
\label{sec:result_compact}

\begin{figure*}
	\includegraphics[width=\textwidth]{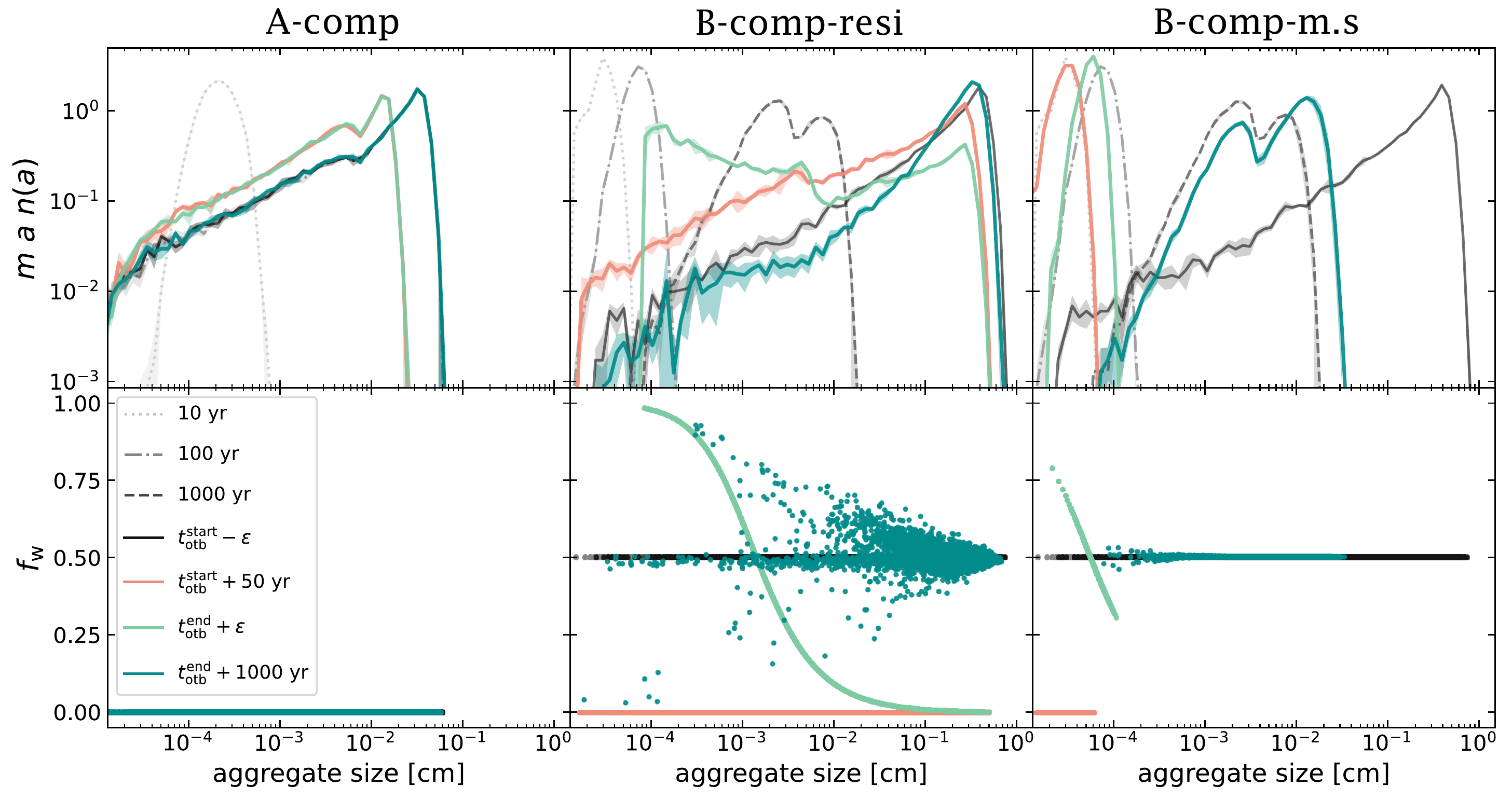}
    \caption{Size distribution function and water fraction of compact aggregates at different times: in the initial quiescent phase (grey shades), during the accretion outburst (red), and after (blue shades).  $t_\mathrm{otb}^\mathrm{start} - \epsilon$ represents the state of the system right before sublimation and $t_\mathrm{otb}^\mathrm{end} + \epsilon$ right after re-condensation. The left panels stand for the solids population in Location A, while the middle and right panels represent the Location B respectively for the resilient and many-seeds model. The area under the size distribution is normalised to $1$ by the total solid mass, being the total rock mass $M_\mathrm{tot}$, or $2M_\mathrm{tot}$ if water ice is present. The shaded areas show the statistical uncertainties, larger for small grains due to the low resolution of the superparticle approach in this part. Each data point in the lower panels represents the properties of a superparticle, itself representing $N_i$ physical particles.}
    \label{fig:compact_BBseeds}
\end{figure*}

\subsubsection{Location A} 
\label{sec:results_compact_A}

We begin with the analysis of \texttt{A-comp} simulating the compact growth of dry particles in location A. When studying dust coagulation, it is common to display the growth of dust aggregates using the size distribution in terms of $m~a~n(a)$, which highlights how the mass is distributed into the population when using logarithmic size bins. Such mass distributions are shown in Fig.~\ref{fig:compact_BBseeds} for different key times, along with the results from the other compact models, respectively \texttt{B-comp-resi} and \texttt{B-comp-m.s}. The distribution of the water mass fraction is also plotted in the bottom panels. For each model, we performed three runs with different random seeds to reduce the statistical noise arising from the Monte Carlo approach. The standard deviation is shown as shaded area on the size distributions.

Location A being located in the very inner disc, the density is high and aggregates collide often, resulting in a short coagulation timescale and a rapid growth. In fact, aggregates grow close to mm-size within $50 \mathrm{~yr}$, in agreement with other coagulation simulations performed in the literature \citep[e.g.][]{brauer2008coagulation}. The size distribution at $100 \mathrm{~yr}$ is characteristic of a coagulation/fragmentation equilibrium, also referred to as collisional equilibrium in this manuscript, where the fragmentation of the large pebbles balances the growth of the fragments. At this stage, the population is in a collisional equilibrium with most of its mass in the upper-end of the distribution, close to the maximum size $a_\mathrm{max}$ whose exact position is determined by disc and dust properties \citep[][]{birnstiel2011dust}. During the outburst, we observe a decrease of the maximum size by a factor ${\approx}3$, before growing back to the pre-outburst state after the event. In fact, $a_\mathrm{max}$ is inversely proportional to the temperature $T$ \citep[][]{birnstiel2011dust}, as an increased temperature results in higher relative velocities, which forces the population to find a new collisional equilibrium corresponding to a scaled-down version of the pre-outburst distribution \citep[see also Fig.6 from ][]{birnstiel2011dust}. The water ice content (bottom-left panel in Fig.~\ref{fig:compact_BBseeds}) remains zero as location A is inside the water snowline at all times.

A similar situation arises in Zone C (see Fig.~\ref{fig:cartoon}), where water always remains in the ice phase. We performed simulations in this region at $25\mathrm{~au}$, however, due to the low surface density and increased coagulation timescale, the outburst was found to be too short to lead to any noticeable changes in the size distribution. We conclude that the only way for an accretion outburst to effectively alter dust aggregates in the outer disc, where $t_\mathrm{coag} \gg \tau_\mathrm{otb}$, is by inducing a compositional change, itself leading to an instantaneous modification of dust structure and properties.

\subsubsection{Location B: Resilient model}

In location B (middle and right column of Fig.~\ref{fig:compact_BBseeds}), the growth is slower and it takes ${\approx}3000\mathrm{~yr}$ for \texttt{B-comp-resi} and \texttt{B-comp-m.s} to reach the fragmentation-limited distribution. Pebbles are more than an order of magnitude larger than in \texttt{A-comp}, which is notably explained by the higher resistance to fragmentation of ice-rich aggregates \citep[ $a_\mathrm{max} \propto v_\mathrm{f}^2$, see][]{birnstiel2009dust}.

At $t_\mathrm{otb}^\mathrm{start}$, aggregates in the resilient model survive water ice sublimation, but still lose $50\%$ of their mass. The fragmentation velocity decreases to the bare rock value, causing the largest surviving pebbles (those close to $a_\mathrm{max}^\mathrm{ice}$) to fragment upon their next collision, efficiently redistributing mass to smaller grains and raising the tail of the size distribution. At $t_\mathrm{otb}^\mathrm{end}$, the temperature decreases again and water re-condenses. Even though pebbles dominate the total mass of the population, it is the small dust grains that dominate the total surface area \citep[e.g.][]{stammler2017redistribution}. As a result, the relative gain in water content is larger for smaller particles, which creates a compositional variation amongst the size distribution, highly diverging from the constant $f_\mathrm{w} = 0.5$ before the outburst. Pebbles slowly regain their water content through collisional mixing with water-rich grains, but, in the meantime, they keep fragmenting efficiently due to their lowered resistance. The water ice distribution has still not fully returned to pre-outburst conditions even after $1000 \mathrm{~yr}$ (dark-blue dots in Fig.~\ref{fig:compact_BBseeds}).

\subsubsection{Location B: Many-seeds model}

In the many-seeds model, the quiescent growth phase is identical to the one in the resilient case, but icy aggregates are assumed to disintegrate as water ice leaves upon heating, effectively resetting the size distribution at $t_\mathrm{otb}^\mathrm{start}$. The size distribution during the outburst is therefore particularly different, as after only $50 \mathrm{~yr}$ dust grains are still in the early-growth phase (red curve in Fig.~\ref{fig:compact_BBseeds}). Having a narrower size distribution, we observe, upon re-condensation, a smaller spread in the water fraction than in the resilient model. $1000\mathrm{~yr}$ after the outburst, the population is still growing and has not reach its collisional equilibrium yet, but the water fraction has bounced back to the initial state.

\subsection{Porous growth}
\label{sec:result_porous}

\begin{figure*}
	\includegraphics[width=\textwidth]{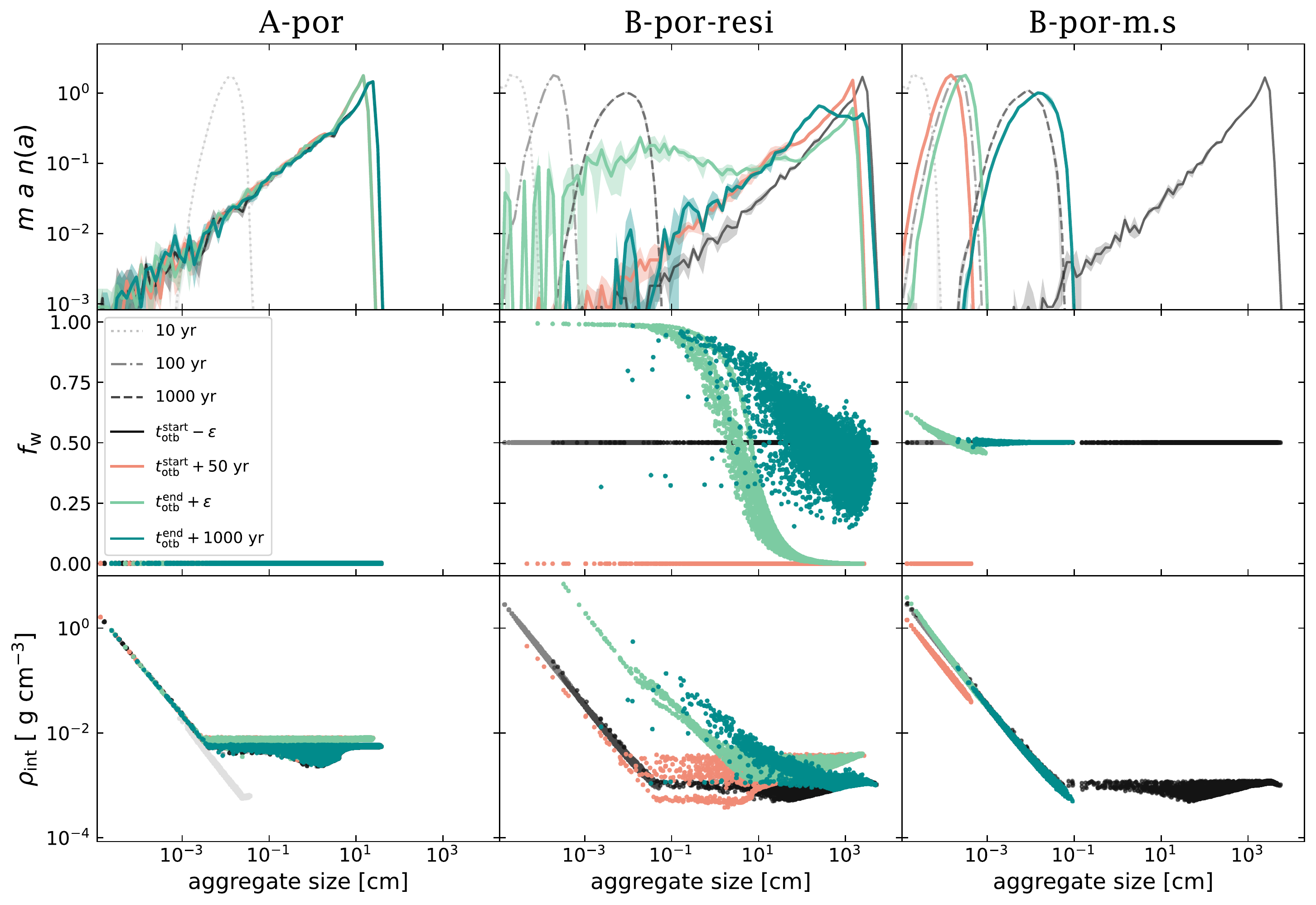}
    \caption{Size distribution function, water fraction, and density evolution of porous aggregates at different times. Same as Fig.~\ref{fig:compact_BBseeds}.}
    \label{fig:porous_ABBseeds}
\end{figure*}

We now discuss the growth of dust aggregates in the porous aggregation. Aggregates porosity is assessed by their internal density, $\rho_\mathrm{int}$, showed in Fig.~\ref{fig:porous_ABBseeds} (bottom panels) at different times for the three models featuring porous growth. Similarly to Fig.~\ref{fig:compact_BBseeds}, we also represent the size distribution (top panels) and water mass fraction (middle panels). 

\subsubsection{Location A}

Similarly to \texttt{A-comp}, aggregates in \texttt{A-por} evolve rapidly and reach the coagulation/fragmentation equilibrium within $50 \mathrm{~yr}$ during the quiescent disc phase. Note that the distribution reaches a maximum size about 2 orders of magnitude larger than the compact case. Several factors contribute to the difference with the compact scenario, as stated in Eq.~\ref{eq:frag_limit_porous}, porous aggregates have a higher resistance to fragmentation. But what mostly influences their larger maximum size arises from their modified aerodynamical behaviour and relative velocities. It can be seen on Fig.~\ref{fig:velocity_A_B}, where identical relative velocities are reached by porous aggregates with much larger sizes \citep[see also Fig.~2 from][]{krijt2015erosion}. The internal density plot displays the different regions introduced in Sect.~\ref{sec:porousaggregation}, with the hit-and-stick regime at small sizes followed by an almost constant phase characteristic of the balance between compression mechanisms and the creation of new voids \citep[][]{okuzumi2012rapid}. The first generation of aggregates (light-grey dots after $10 \mathrm{~yr}$ in Fig.~\ref{fig:porous_ABBseeds}) is more porous, it is then compressed at higher sizes by gas ram pressure before fragmenting into equal or higher density fragment (see fragmentation prescription in Sect.~\ref{sec:fragmentation}), which explains why they do not reappear in later stages. Fragmentation also prevents the formation of even larger solids which would become denser due to self-gravity compression \citep[][]{kataoka2013fluffy}.

As in the compact case, the outburst leads to a temporary decrease in the maximum size. The amplitude of the variation is identical, as changing disc conditions vary independently of the aggregation model. We also see a slight increase in the aggregate internal density, which is related to an enhanced compression by collisional restructuring and gas ram pressure. After the outburst, the system quickly comes back to the pre-outburst state.

\subsubsection{Location B: Resilient model}

Porous aggregates in location B grow until establishing their collisional equilibrium, once again corresponding to superior maximum size as location A due to the presence of water ice. The internal density follows the aforementioned porous model, although the lower gas ram pressure and larger rolling energy of icy monomers leads to more porous aggregates.

At $t_\mathrm{otb}^\mathrm{start}$, resilient aggregates survive the sublimation of water and lose $50\%$ of their mass. The impact on the internal density depends on their size. Aggregates below ${\approx}10 \mathrm{~cm}$ retain their size, and the loss of mass then results in a decrease in the internal density. For larger aggregates, however, the story is more complex. Here, the lowered rolling energy $E_\mathrm{roll}^\mathrm{rock}$ and increased gas ram pressure leads to an additional compression, and the internal density actually increases. These denser aggregates then fragment and generate grains of equal or higher density (see Sect.~\ref{sec:fragmentation}), which ends up creating a broader density distribution (red dots in the bottom panel in Fig. \ref{fig:porous_ABBseeds}).

The re-condensation of water follows the same trend as in \texttt{B-comp-resi}, with few differences arising from the impact of the porosity on the aggregates' surface area. The central slope appears broader (light-blue dots in the central panel in Fig.~\ref{fig:porous_ABBseeds}), due to the similar trend in the internal density distribution. We also notice that grains in the hit-and-stick regime take an even larger fraction of water ice, all ending with similar and very high ice contents. The reason for this is that in this hit-and-stick phase, the fractal dimension $D_\mathrm{f} \approx 2$, leading to $m \propto a^2$ and a surface area per mass unit that is similar for each aggregate. They thus receive an amount of water leading to a similar fraction $f_\mathrm{w}$ as the others in the hit-and-stick regime. We see that water re-condensation gives rise to intermediate-sized aggregates filled with water ice and displaying large internal density. After $1000 \mathrm{~yr}$, the population still did not reach the pre-outburst state.

\subsubsection{Location B: Many-seeds model}

In \texttt{B-por-m.s}, all aggregates are still in the hit-and-stick regime when the outburst ends, leading to even narrower water mass distribution than in \texttt{B-comp-m.s}. Despite leading to dramatic size alteration, we see that aggregates following the many-seeds response are characterised with narrower water distribution than resilient ones in both aggregation model. However it may differ for longer outbursts, where dust aggregates would have longer time to grow before re-condensation occurs, especially if they reach a different stage of their porous evolution. The system then keeps growing, and $1000 \mathrm{~yr}$ after the outburst it is still growing in the hit-and-stick regime while having almost fully recover the initial water distribution at $f_\mathrm{w}=0.5$.

\subsection{Mass-weighted size and Stokes number}
\label{sec:maximumsize}

\begin{figure*}
	\includegraphics[width=\textwidth]{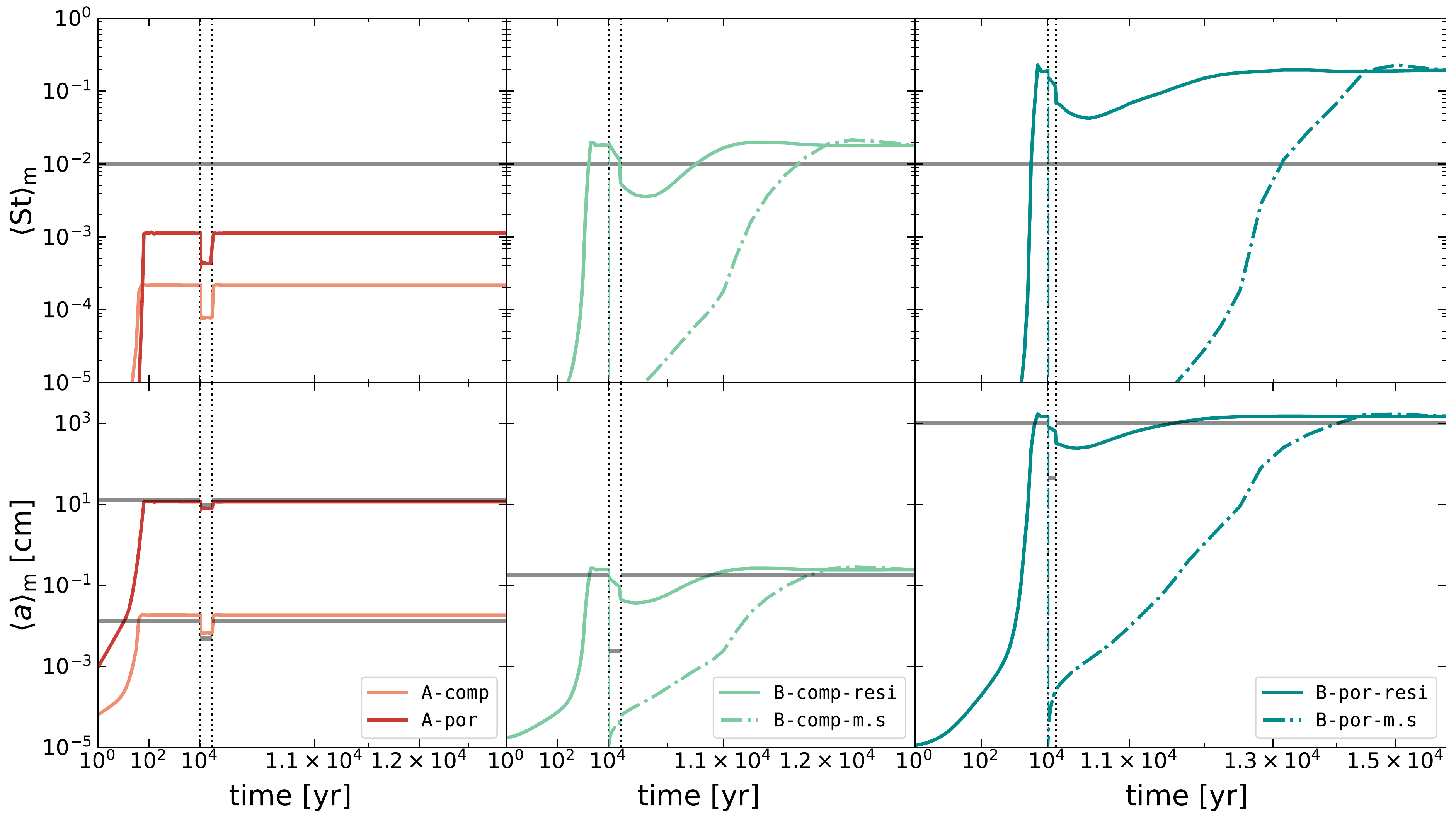}
    \caption{Mass-weighted average size $\langle a \rangle_\mathrm{m}$ and Stokes number $\langle St \rangle_\mathrm{m}$ vs. time. The vertical lines denote respectively $t_\mathrm{otb}^\mathrm{start}$ and $t_\mathrm{otb}^\mathrm{end}$. The horizontal grey lines show: the typical value needed to trigger the streaming instability following \citet[][]{li2021thresholds} (top), and the theoretical position of the collisional equilibrium for each model in the quiescent and outburst phase (bottom). The logarithmic scale is modified at $t_\mathrm{otb}^\mathrm{start}$ to better discern the variations caused by the event. The population being well resolved at large sizes (see Fig.~\ref{fig:compact_BBseeds} and Fig.~\ref{fig:porous_ABBseeds}), we do not include the statistical noise calculated from independent runs.}
    \label{fig:max_size}
\end{figure*}

In this section, we investigate the temporal evolution of the mass-weighted average size $\langle a \rangle_\mathrm{m}$ and Stokes number $\langle \mathrm{St} \rangle_\mathrm{m}$ (calculated from Eq.~\ref{eq:stokesnumber}), shown in Fig.~\ref{fig:max_size}. Because in the coagulation/fragmentation equilibrium most of the mass is located close to the maximum size (see Fig.~\ref{fig:compact_BBseeds} and Fig.~\ref{fig:porous_ABBseeds}), these quantities are a good indicator of the properties of the largest aggregates. They are also helpful in determining when the population enters a collisional equilibrium, as a constant size distribution would lead to a constant $\langle a \rangle_\mathrm{m}$ and $\langle St \rangle_\mathrm{m}$. In the bottom panels of Fig.~\ref{fig:max_size}, horizontal grey lines indicate estimates of $\langle a \rangle_\mathrm{m}$ in the collisional equilibrium expected in the quiescent and outburst phase of the different models. It is found by solving for the size at which equal aggregates collide at $v_\mathrm{f}-\delta v_\mathrm{f}$ (see Appendix \ref{sec:appendixA}), where collisions begin to lead to mass loss.

During the quiescent phase, $\langle a \rangle_\mathrm{m}$ increases with time for each model until reaching the plateau characteristic of their respective coagulation/fragmentation equilibrium. Porous aggregates grow more rapidly but to greater sizes, so that the time it takes to reach the equilibrium is similar to the compact case. At $t_\mathrm{otb}^\mathrm{start}$, high collision rates in location A leads to a rapid adjustment, and the new scaled-down equilibrium is reached within a few years. After $t_\mathrm{otb}^\mathrm{end}$, the population recovers to the pre-outburst distribution also within a few years in both aggregation models. $\langle a \rangle_\mathrm{m}$ thus closely matches the theoretical prediction before, during, and after the outburst, meaning that the population is in collisional equilibrium at almost all times.

The picture is more complex in location B, where the lower surface density leads to smaller collision rates. The compact (middle panel) and porous (right panel) populations are far from reaching the new equilibrium within the outburst duration. In the resilient model, we will find pebbles in our outbursting disc that are too large for their rocky composition. Interestingly, the lowest $\langle a \rangle_\mathrm{m}$ is reached a few hundreds of years after the event, when the largest pebbles are still deprecated in water ice and effectively fragmenting (see Fig.~\ref{fig:compact_BBseeds} and Fig.~\ref{fig:porous_ABBseeds}). After the water ice is redistributed through collisional mixing, $\langle a \rangle_\mathrm{m}$ returns to pre-outburst value and the collisional equilibrium is re-established.

In the many-seeds model, aggregates disintegrate to monomers, and, just like in the resilient case, they do not have the time to reach the new equilibrium within the outburst duration. Although in the many-seeds case, aggregates are now below that theoretical value. After the outburst, they take longer to re-establish the collisional equilibrium than the resilient models. We conclude that in our simulations in location B, the use of the collisional equilibrium distribution is never appropriate to describe the dust population during the outburst, and it remains so after the outburst for a duration depending on the model (up to $4500\mathrm{~yr}$ at $5 \mathrm{~au}$ in \texttt{B-por-m.s}).

The Stokes number $\langle \mathrm{St} \rangle_\mathrm{m}$ is represented on the top panels and displays a similar evolution as $\langle a \rangle_\mathrm{m}$. We include a horizontal line at $\mathrm{St}=0.01$, which indicates the minimum value necessary to trigger the streaming instability at dust to gas ratios close to $10^{-2}$ \citep[][]{li2021thresholds}. In location A, we remain below that limit at all times. In location B, the efficient fragmentation phase of pebbles in \texttt{B-comp-resi} also results in the Stokes number to drop below $0.01$ for a duration of almost $1000 \mathrm{~yr}$. In the many-seeds model, it takes significantly longer to re-grow pebbles above $\mathrm{St}=0.01$, about $1500$ and $3000 \mathrm{~yr}$ for \texttt{B-comp-m.s} and \texttt{B-por-m.s}, respectively. 

We performed additional coagulation simulations accounting for different values of the turbulence strength, respectively $\alpha=10^{-4}$ and $10^{-2}$. Lower turbulence pushes all models further from reaching the outburst collisional equilibrium within $\tau_\mathrm{otb}$, as aggregates in the resilient model fragment less often due to lower turbulent velocity and collision rates, while in the many-seeds grains have to re-grow to larger size \citep[ $a_\mathrm{max} \propto \alpha^{-1}$, see][]{birnstiel2009dust} under similar lowered collision rates. For higher turbulence strength, the reverse situation occurs bringing all models closer to reaching the new equilibrium within the outburst duration. In such a case, the post-outburst phase could be similar for both sublimation models (see also fast and intermediate adjustments in Fig.~\ref{fig:cartoon_recovery}).

\section{Observational signatures}
\label{sec:opacity_model}

In this section, we investigate how the alteration of dust properties affects their observational signatures during and after the accretion outburst. We convert the results of our coagulation simulations into absorption opacities using the \texttt{DSHARP-OPAC} package from \citet[]{birnstiel2018disk}. Given the dust composition we adopted (Table \ref{tab:dustcomposition}), the optical constants are taken from \citet{warren2008optical} for water ice, \citet{draine2003interstellar} for astronomical silicates, and \citet{henning1996dust} for troilite and refractory organics. We compute the mixed dielectric function using the Bruggeman effective medium theory, that is applicable when the different materials are homogeneously mixed with no dominant medium. Note that directly after the outburst, this may not be the case for highly water-rich grains, but simulations show that these small particles are rapidly mixed with the rest of the population. In the porous aggregation model, we additionally use the Maxwell-Garnet rule to account for the voids arising from the porous structure \citep[]{voshchinnikov2005modelling, kataoka2014opacity}. Opacities of individual aggregates are computed using Mie theory, considering their unique size, composition and porosity, and combined into the total absorption opacity $\kappa_\lambda^\mathrm{abs,tot}$ by summing over the distributions returned by the coagulation calculations\footnote{Following Equation (6) from \citet[][]{birnstiel2018disk}, the denominator equals to the total solid mass of the population, being the total rock mass $M_\mathrm{tot}$, or $2M_\mathrm{tot}$ if water ice is present.}. Note that for computational optimization, we do not calculate the opacity of each individual superparticle, but rather group particles with similar mass, size, and composition. Fig.~\ref{fig:totalopacity} displays the total absorption opacity of the population at different key times of the simulation. Being sensitive to the entire dust distribution, the absorption opacity could be altered by the lower resolution of the superparticle approach towards small grains, which displays important statistical noises after re-condensation in the resilient models (see Fig.~\ref{fig:compact_BBseeds} and Fig.~\ref{fig:porous_ABBseeds}). For \texttt{B-comp-resi} and \texttt{B-por-resi} (central panels), we thus performed opacity computations for three independent runs and include the statistical uncertainties as shaded areas. We see that they are barely noticeable, hence have a negligible impact on our results.

\begin{figure}
	\includegraphics[width=\columnwidth]{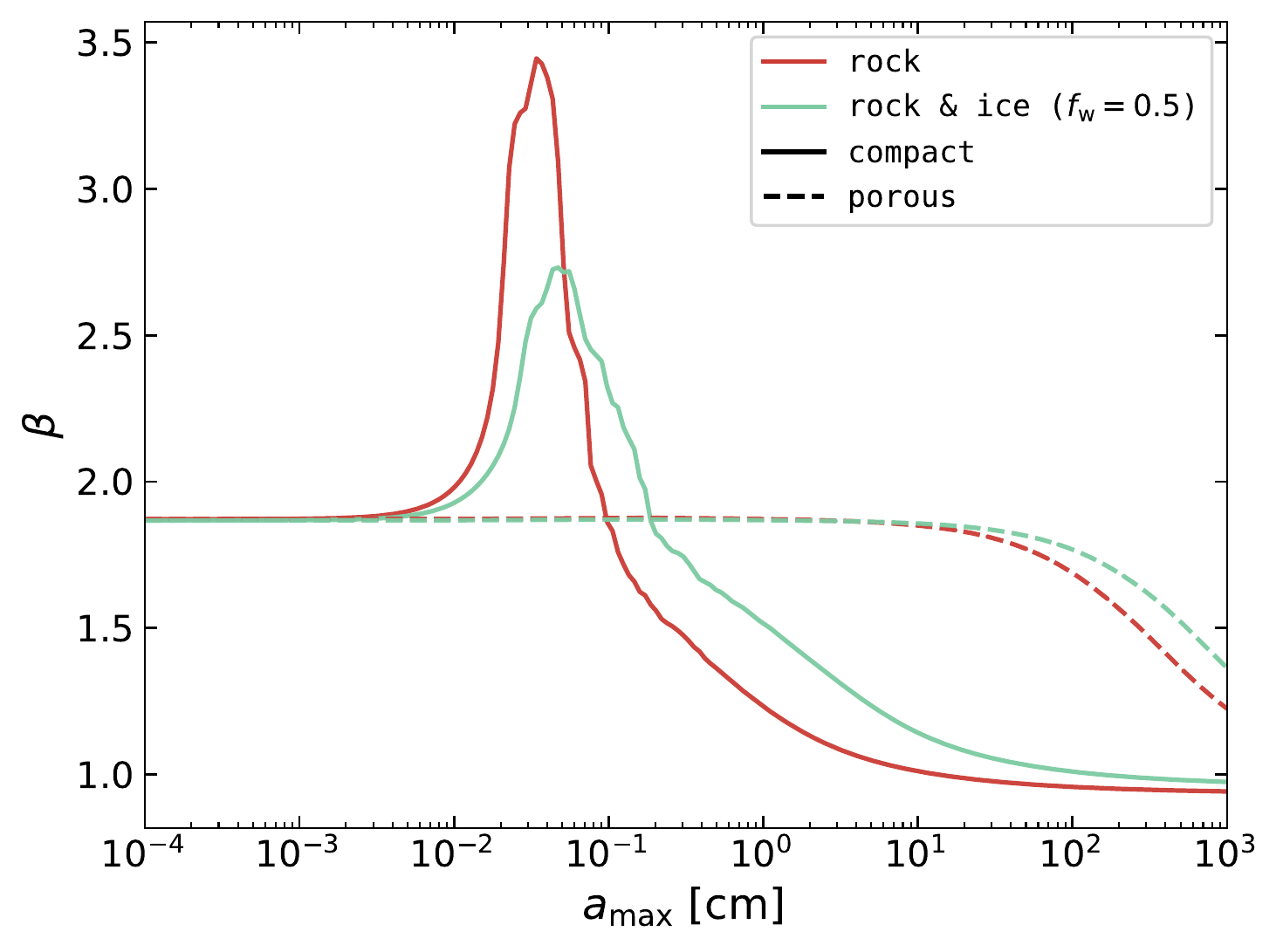}
    \caption{$\beta$ as a function of maximum particle size, assuming a power-law size distribution $n(a) \propto a^{-3.5}$ from $10^{-4} \mathrm{~cm}$ to $a_\mathrm{max}$ \citep[more details in Sect.3 from ][]{birnstiel2018disk}. The two compositions correspond to the mixtures in Table \ref{tab:dustcomposition}. For the porous case, we use $\rho_\mathrm{int}=10^{-2}  ~\mathrm{g~cm^{-3}}$, similarly to what we obtain in our simulations (Fig.~\ref{fig:porous_ABBseeds}).}
    \label{fig:beta_vs_maxsize}
\end{figure}

\begin{figure*}
	\includegraphics[width=\textwidth]{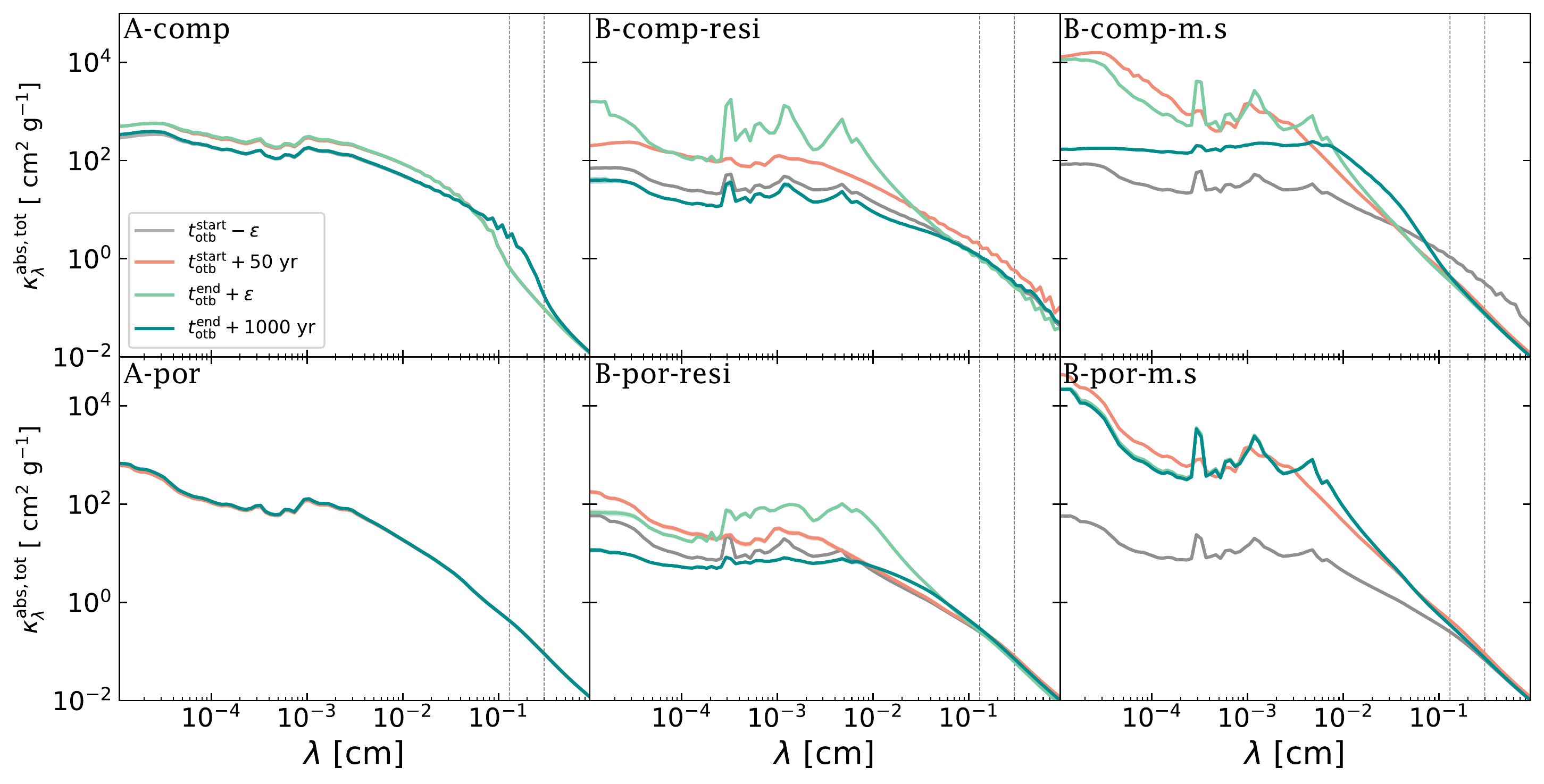}
    \caption{Absorption opacity of the grains distribution at different key times for the different locations and models, following the same color code as Fig.~\ref{fig:compact_BBseeds} and Fig.~\ref{fig:porous_ABBseeds}. $t_\mathrm{otb}^\mathrm{start} - \epsilon$ represents the state of the population right before sublimation and $t_\mathrm{otb}^\mathrm{end} + \epsilon$ directly after re-condensation. The left panels stand for the population in Location A, while the middle and right panels represent the Location B respectively for the resilient and many-seeds model. The wavelengths used to compute $\beta$ are denoted with vertical lines. We included the statistical uncertainties on the middle panels similarly to Fig.~\ref{fig:compact_BBseeds} and Fig.~\ref{fig:porous_ABBseeds}, as the absorption opacity is sensitive to the entire size distribution which was partly unresolved at small sizes for these two models after the outburst.}
    \label{fig:totalopacity}
\end{figure*}

\begin{figure*}
	\includegraphics[width=\textwidth]{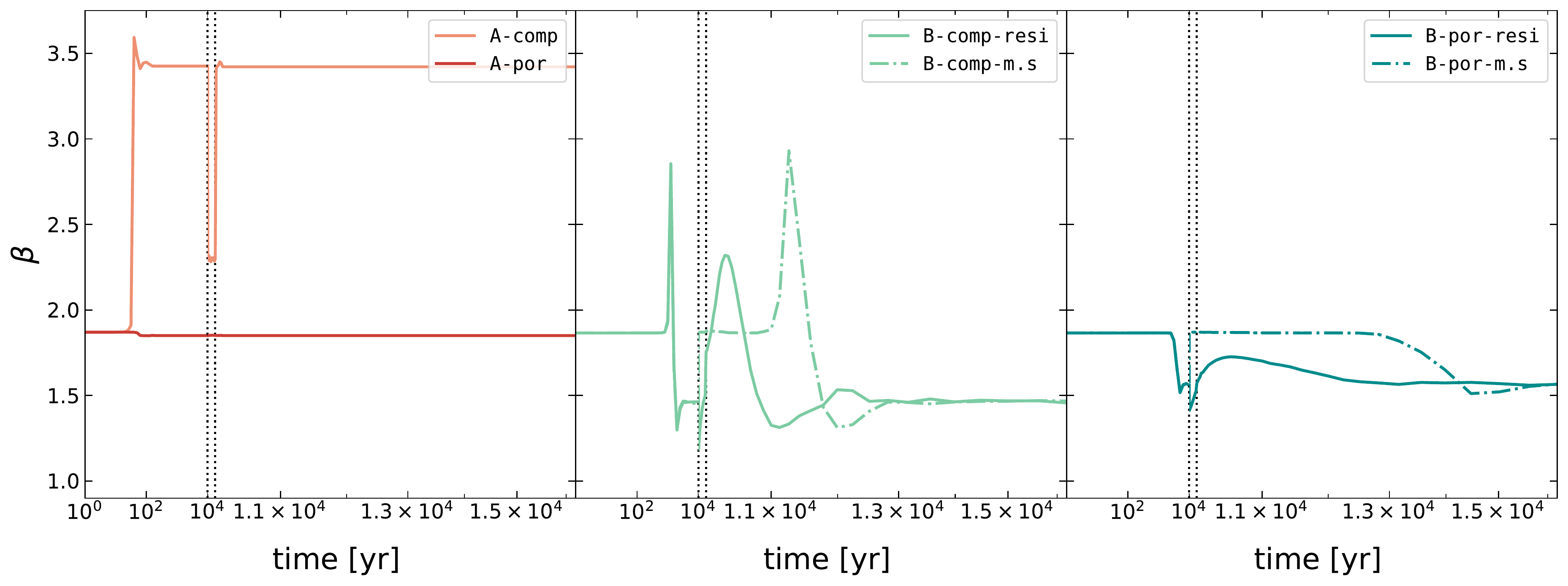}
    \caption{$\beta$ vs. time for the different locations and models.The vertical lines denote respectively $t_\mathrm{otb}^\mathrm{start}$ and $t_\mathrm{otb}^\mathrm{end}$. The logarithmic scale is reset at $t_\mathrm{otb}^\mathrm{start}$ to better discern the features of the outburst, similarly to Fig.~\ref{fig:max_size}. The population being well resolved at millimetre wavelengths (see Fig.~\ref{fig:totalopacity}), we do not include the statistical noise of  $\beta$ calculated from three independent runs.}
    \label{fig:beta}
\end{figure*}

We can also compute $\beta$, the spectral index of the dust opacity $\kappa_{\nu} \propto \nu^{\beta}$, which is widely used in the literature to trace the properties of millimetre-sized particles in protoplanetary discs \citep[][]{beckwith1990survey}. In this paper, $\beta$ is computed using $1.3$ and $3\mathrm{~mm}$, corresponding respectively to Band 6 and 3 of The Atacama Large Millimeter/submillimeter Array (ALMA), as it is the most powerful tool to study protoplanetary discs and probe particle properties near the disc midplane \citep[][]{andrews2020observations}. The spectral index is then given by
\begin{equation}
\beta = - \dfrac{\mathrm{log}(\kappa_{3\mathrm{mm}}/\kappa_{1.3\mathrm{mm}})}{\mathrm{log}(\nu_{3\mathrm{mm}}/\nu_{1.3\mathrm{mm}})}.
\end{equation}
To illustrate how dust coagulation can impact the opacity index, we first represent $\beta$ as a function of the maximum particle size on Fig.~\ref{fig:beta_vs_maxsize}, which is calculated using a simplified power-law distribution with a cut-off at $a_\mathrm{max}$ and a slope $q=3.5$. This plot offers a comprehensive overview of the significant impact that particle size, composition, and porosity have on the opacity index. The outburst affecting each property (see Fig.~\ref{fig:compact_BBseeds} and Fig.~\ref{fig:porous_ABBseeds}), it will result in temporal variations of the opacity index, which we represent on Fig.~\ref{fig:beta}, computed from the results of our coagulation simulations.

We begin our analysis with Location A. Early on, $a \ll 1 \mathrm{~mm}$ and $\beta$ is constant, close to $1.7$ as for dust grains in the ISM \citep[][]{finkbeiner1999extrapolation}. It then diverges depending on the aggregation model. For compact growth, $\beta$ peaks when the size distribution approaches $a \approx \lambda/2\pi$ and resonances amplify the opacity (Fig.~\ref{fig:beta_vs_maxsize}). The collisional equilibrium distribution being close to the resonance, $\beta$ remains relatively large. The outburst leads to a strong variation in $\beta$ due to the fragmentation of aggregates in the resonance size range (see Fig.~\ref{fig:compact_BBseeds}). It can also be seen on the total absorption opacity, that is higher in the outburst phase, except above $5 \times 10^{-2} \mathrm{~cm}$ due to the redistribution of mass in smaller fragments. For porous aggregates, the opacity is sensitive to the mass-to-area ratio \citep[][]{kataoka2014opacity}, which mostly varies when the population reaches efficient compression mechanisms early in its evolution (see light-grey dots on Fig.~\ref{fig:porous_ABBseeds}). At millimetre wavelengths, the resonant amplifications of the opacity is damped (Fig.~\ref{fig:beta_vs_maxsize}), and $\beta$ remains mostly constant throughout the quiescent and outbursting phases.

In location B, for the compact growth, we also observe the constant $\beta$ followed by a resonant amplifications. Then, $\beta$ decreases as the population grows above millimetre sizes to settle at the coagulation/fragmentation equilibrium. When the outburst starts, the evolution diverges depending on the sublimation model. In the many-seeds case (\texttt{B-comp-m.s}), aggregates fall apart, and $\beta \approx 1.7$. Aggregates recover with time and the resonant amplifications is observed again after $1200 \mathrm{~yr}$, with $\beta \approx 2.9$. In Fig.~\ref{fig:totalopacity}, we can see the water features reappearing strongly after the outburst due to the large amount of small icy particles. It dampens with collisional mixing before recovering to the pre-outburst spectrum. For resilient aggregates (\texttt{B-comp-resi}), $\beta$ decreases sharply when water ice sublimates. It increases during the outburst due to the enhanced fragmentation of dry aggregates above millimetre sizes. Fig.~\ref{fig:totalopacity} also shows this behaviour with an overall increase in the total absorption opacity, along with a disappearance of water features. After the outburst, water features reappear and $\beta$ peaks at about $2.3$ a few hundreds of years after the event, when pebbles stop fragmenting. Similarly to the pre-outburst, $\beta$ slightly re-increases when the collisional equilibrium is found, although it takes longer than for the many-seeds case. The recovery seems longer in Fig.~\ref{fig:beta} than in Fig.~\ref{fig:max_size}, as $\beta$ depends the properties of the entire distribution which take more time than only recovering $\langle a \rangle_\mathrm{m}$.

\begin{figure}
	\includegraphics[width=\columnwidth]{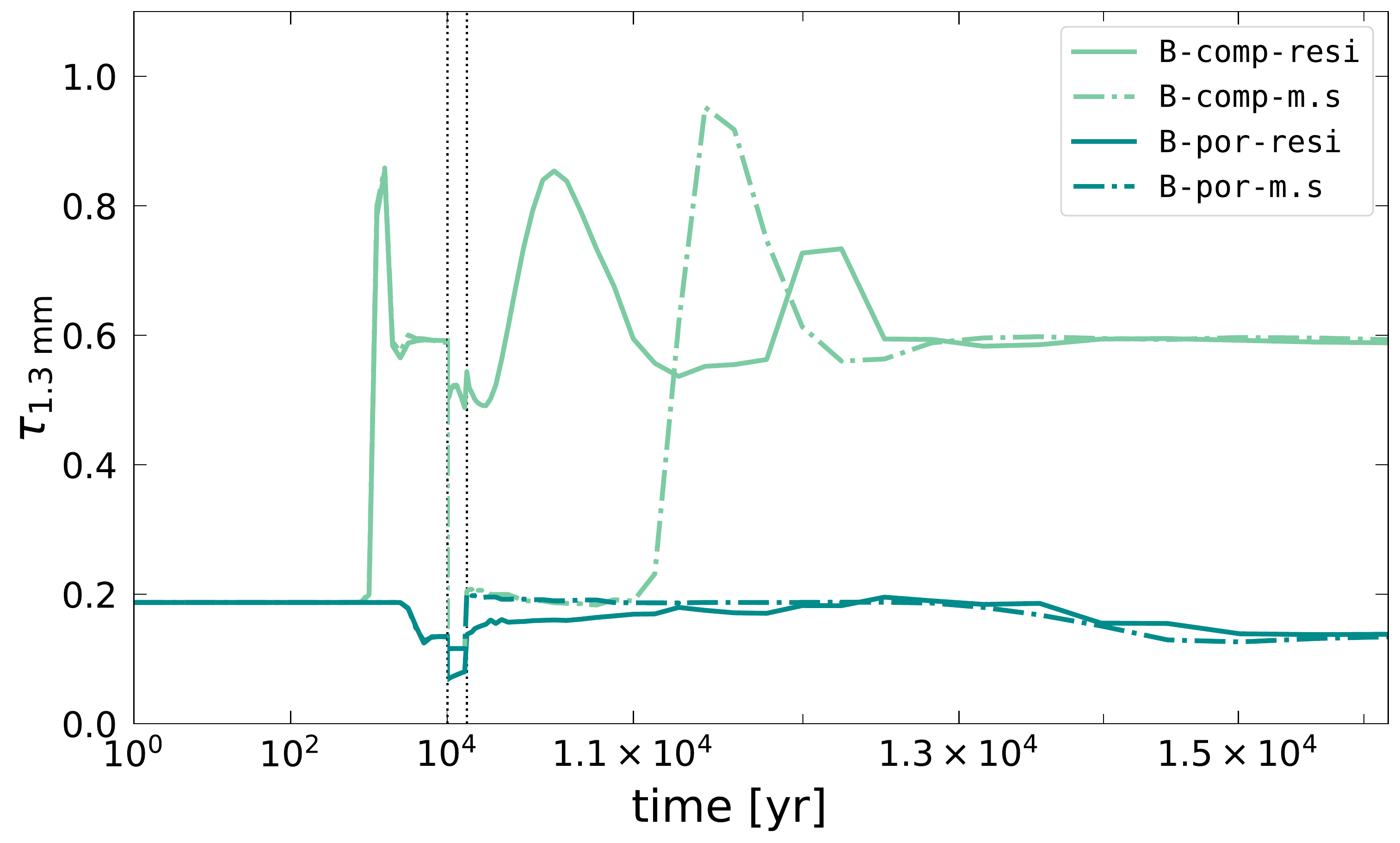}
    \caption{Optical depth at $1.3 \mathrm{~mm}$ vs. time for the different models in location B. The vertical lines denote respectively $t_\mathrm{otb}^\mathrm{start}$ and $t_\mathrm{otb}^\mathrm{end}$. The logarithmic scale is reset at $t_\mathrm{otb}^\mathrm{start}$ (see also Fig.~\ref{fig:max_size}). The population being well resolved at millimetre wavelengths (see Fig.~\ref{fig:totalopacity}), we do not include the statistical noise from independent runs.}
    \label{fig:opt_depth}
\end{figure}

Concerning the porous model, the quiescent phase behaves similarly to location A (\texttt{A-por}) with the resonant amplifications being damped. However, aggregates in B reach larger sizes (${>}10^2 \mathrm{~cm}$) where $\beta$ starts to decrease (Fig.~\ref{fig:beta_vs_maxsize}). For \texttt{B-por-m.s}, the evolution of $\beta$ is reset. After $1000 \mathrm{~yr}$, the population is still in the hit-and-stick regime with a constant mass-to-area ratio, which explains why the post-outburst opacity spectra are identical. For \texttt{B-por-resi}, $\beta$ does not exhibit major variations, except for a slight increase during the efficient fragmentation phase during the outburst. Note that for the four models in location B, we represented in Fig.~\ref{fig:opt_depth} the temporal evolution of the optical depth at $1.3 \mathrm{~mm}$, calculated as $\tau_{1.3 \mathrm{~mm}} = \kappa_{1.3 \mathrm{~mm}}^\mathrm{abs, tot} \Sigma_\mathrm{d}(5 \mathrm{~au})$, i.e. assuming our midplane simulations represent well the entire disc column. Note that the solid surface density is doubled when water is in ice form, as the dust-to-ice ratio is unity (Sect.~\ref{sec:watercontent}). We see that the optical depth is below unity throughout the simulations, meaning that the emissions are optically thin and $\beta$ effectively connects to the dust size distribution and properties \citep[][]{testi2014dust}.

In the end, we can see that the outburst induces a wide range of observable signatures, highly dependant on the size distribution, aggregation model, and response to sublimation. The recovery of compact aggregates in zone B leads to particularly strong variations in $\beta$, even long after the outburst ended. Porous aggregates, however, lack strong variations at millimetre wavelengths due to the absence of resonant amplifications. We will discuss these features in Sect.~\ref{sec:pastoutburst}. Note that in our simulations, the signals appear in the aftermath of the outburst, but in case of a longer outburst or shorter coagulation timescale, they may even emerge during the event.


\section{Discussion}
\label{sec:discussion}

\subsection{Outburst and post-outburst adjustments}
\label{sec:postoutburstrecovery}

\begin{figure*}
	\includegraphics[width=\textwidth]{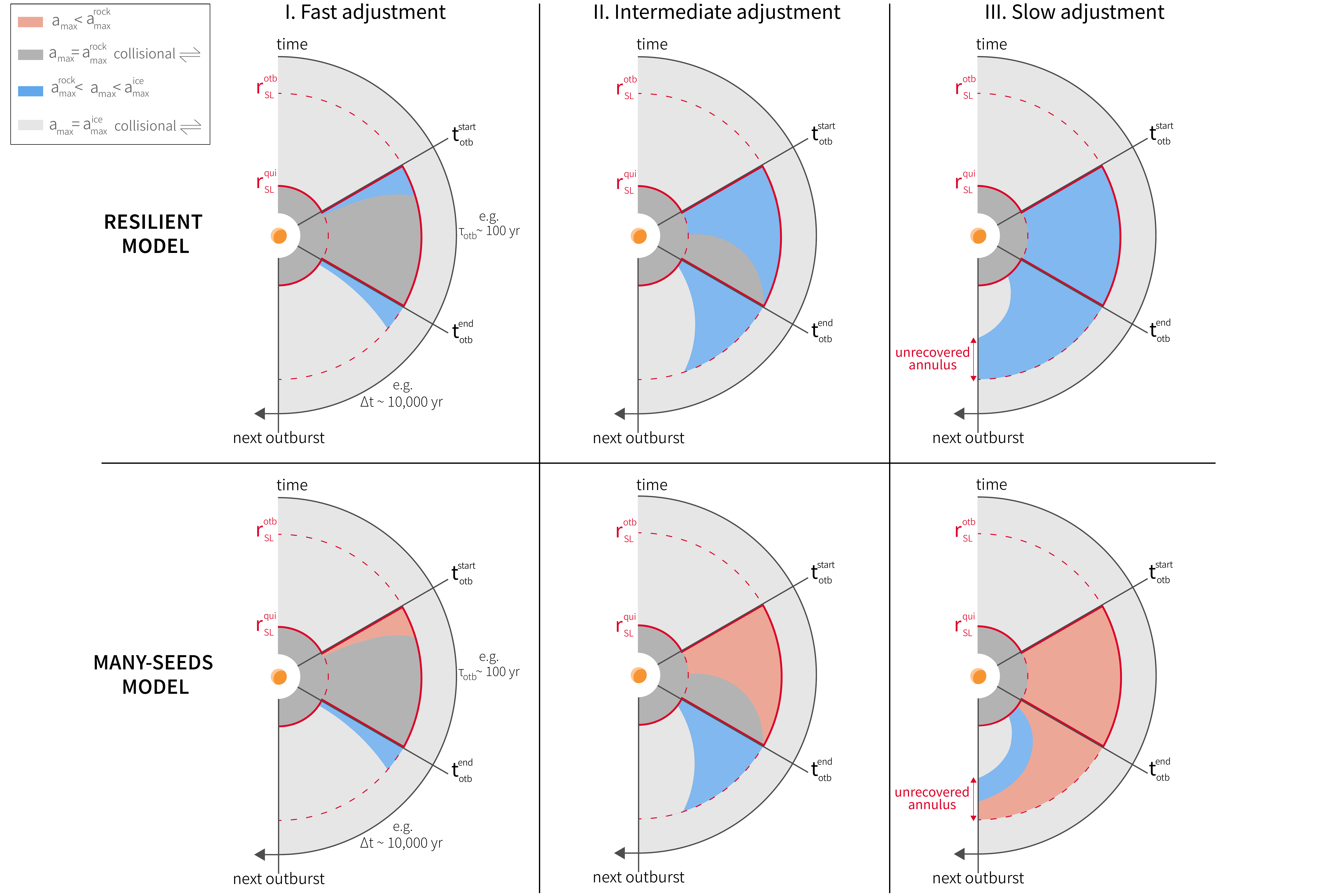}
    \caption{Schematic illustrating the adjustment of the dust size distribution in the radial direction during and after the outburst, for the resilient and many-seeds models. (1) Fast adjustment: the size distribution adjusts rapidly to the outburst conditions and recovers rapidly after, so that dust grains are in collisional equilibrium for most of the outburst duration and quiescent phase. (2) Intermediate adjustment: The outer part of the disc responds less rapidly, hence it is not in equilibrium during most of the outburst duration and quiescent phase. The dust population entirely recovers before the next event. (3) Slow adjustment : All solids between $r_\mathrm{SL}^\mathrm{qui}$ and $r_\mathrm{SL}^\mathrm{otb}$ do not reach the collisional equilibrium during the outburst. Parts of the outer disc do not recover before the next event, leading to a non-recovery annulus whose width depends on the coagulation physics and outburst properties.}
    \label{fig:cartoon_recovery}
\end{figure*}

As we saw in Sect.~\ref{sec:results}, accretion outbursts modify the disc temperature and the properties of dust particles, and a certain time span is needed for the population to respond to these changes, and to recover to the initial (quiescent) equilibrium after the outburst has passed. In this section, we investigate the adjustment timescale and compare it to the outburst duration and rate. We will only focus the discussion on zone B, as zone A is characterised by relatively fast adjustment (Fig.~\ref{fig:max_size}) thanks to high collision rates in the inner disc and smaller size variation (\ref{sec:results_compact_A}).

Depending on the coagulation physics and outburst properties, we summarize in Fig.~\ref{fig:cartoon_recovery} three adjustment cases. I) In the fast adjustment scenario, the dust population in zone B adapts rapidly to the scaled-down bare rock distribution during the outburst, and recovers likewise after the event. Solids thus spend most of their evolution in the corresponding collisional equilibrium. This situation may arise for example due to intrinsic high solid density in the disc, strong turbulence (see Sect.~\ref{sec:maximumsize}), or due to a weaker outburst amplitude (e.g. EXor-type accretion events) keeping the excited snowline relatively close to the host star. 

II) In the intermediate adjustment scenario, the dust population has the time to find the new collisional equilibrium during the outburst, but still spends a large fraction of the event out of equilibrium. The dust content of zone B recovers to the quiescent equilibrium before the occurrence of the next outburst. The complete recovery is even more likely to arise in older Class II discs, for which the time span between outbursts is longer \citep[$\Delta t_\mathrm{otb}\approx 10^5$ yr,][]{contreras2019determining}. 

In these two first cases, the difference between the resilient and many-seeds model is only visible during the outburst, for a given temporal fraction where the largest aggregates are respectively above or below the bare rock maximum size $a_\mathrm{max}^\mathrm{rock}$. As a consequence, unlike in our simulations, most features in $\beta$ would only be visible during the outburst phase, and the re-condensation would behave similarly in both situations given they end the outburst with similar size distributions. These cases also illustrate the model of \citet[][]{schoonenberg2017pebbles}, where the re-coagulation of silicates, after they fell apart (many-seeds model) created a visible structure at $42 \mathrm{~au}$ in the outbursting disc V883 Ori. 

III) Finally, in the slow adjustment scenario, the dust population through zone B is out of equilibrium for the entirety of the outburst duration and for a significant fraction of the quiescent phase, before the occurrence of a next outburst. A part of the disc may even not recover at all, leading to an unrecovered annulus within which the dust population is perpetually out of equilibrium. Our local simulations in location B at $5\mathrm{~au}$ fall between cases II and III: the dust population does not adjust to $a_\mathrm{max}^\mathrm{rock}$ within the outburst duration, but does recover on timescale shorter than $10^4 \mathrm{~yr}$, being the potential next event assuming a constant rate.

If the resilient model applies, an intermediate or slow adjustment may provide an explanation for the recent observations of \citet[][]{liu2021millimeter} of dry millimetre-sized pebbles inside the excited water snowline of the outbursting disc FU Ori. They explain their presence by invoking a higher resistance of bare rocks towards fragmentation than previously thought, but instead, we suggest that these large dry pebbles may simply not have had enough time to experience enough fragmentation collisions (Fig.~\ref{fig:max_size}). It is interesting to notice how different models (resilient and many-seeds) can provide satisfying explanations in two different discs (resp. FU Ori and V883 Ori). Looking at a larger sample of discs, outbursting systems could then provide laboratories for exploring further the behaviour of dust aggregates to sublimation, which is a key aspect in planetesimal formation scenarios at the water snowline \citep[e.g.][]{schoonenberg2017planetesimal}. 

Accretion outbursts being probably frequent and widespread in most forming systems \citep[][]{dunham2012resolving, audard2014episodic}, understanding the recovery process of dust grains after the outburst is of crucial importance to comprehend observed discs, and further constrain outburst properties. We define the recovery timescale $t_\mathrm{rec}$ as the time needed for $\langle a \rangle_\mathrm{m}$ to grow from its value when the outburst ends back to the quiescent collisional equilibrium. At $5\mathrm{~au}$, in the resilient model, Fig.~\ref{fig:max_size} shows it takes approximately $t_\mathrm{rec} \approx 1000\mathrm{~yr}$ for \texttt{B-comp-resi} and $2000\mathrm{~yr}$ for \texttt{B-por-resi} to recover. In the many-seeds model, it takes $2000\mathrm{~yr}$ for \texttt{B-comp-m.s} and $4500\mathrm{~yr}$ for \texttt{B-por-m.s}, longer than in the resilient case. We note that these values are sensitive to our disc and outburst models. As previously mentioned, a longer outburst would give more time to many-seeds aggregates to grow hence lower $t_\mathrm{rec}$, unlike the resilient case where a longer fragmentation phase would decrease the maximum size thus increase $t_\mathrm{rec}$. 

Based on our definition, we can express the recovery timescale\footnote{This expression assumes that the dominant source for relative velocity is turbulence, and considers also the reduced scale-height of dust grains with $St > \alpha$ resulting from vertical settling. In that specific case, the final growth timescale does not depend on the turbulence $\alpha$. While our simulations do not include vertical settling (Sect.\ref{sec:dustdynamics}), we opt for using this expression for the growth timescale to translate our results to the outer disc, where settling may be more pervasive.} as \citep[][]{birnstiel2011dust}
\begin{equation}
    t_\mathrm{rec} \approx \dfrac{1}{\delta_\mathrm{d2g} \Omega} \ln \bigg( \dfrac{\langle a \rangle_\mathrm{m}(t_\mathrm{otb}^\mathrm{start})}{\langle a \rangle_\mathrm{m}(t_\mathrm{otb}^\mathrm{end})} \bigg),
    %\ln{(s_\mathrm{max}/s_\mathrm{otb,end})}.
\label{eq:recovery}
\end{equation}
which we generalise to any heliocentric distance $r$ using our results at $5\mathrm{~au}$ as a point of reference,

\begin{equation}
    t_\mathrm{rec}(r) \approx t_\mathrm{rec}(5\mathrm{au}) \dfrac{\delta_\mathrm{d2g}(5\mathrm{au})}{\delta_\mathrm{d2g}(r)} \bigg( \dfrac{r}{5\mathrm{au}} \bigg)^{3/2} \dfrac{ \ln \bigg( \dfrac{\langle a \rangle_\mathrm{m}(t_\mathrm{otb}^\mathrm{start}, r)}{\langle a \rangle_\mathrm{m}(t_\mathrm{otb}^\mathrm{end}, r)} \bigg)}{ \ln \bigg( \dfrac{\langle a \rangle_\mathrm{m}(t_\mathrm{otb}^\mathrm{start}, 5\mathrm{au})}{\langle a \rangle_\mathrm{m}(t_\mathrm{otb}^\mathrm{end}, 5\mathrm{au})} \bigg)}.
\end{equation}
Considering the dust-to-gas ratio to be constant throughout the disc, and neglecting the last term as the slight variation in the size ratio within the natural logarithm would only impact $t_\mathrm{rec}$ by a few factors, we find 
\begin{equation}
    t_\mathrm{rec}(r) \approx t_\mathrm{rec}(5\mathrm{au}) \bigg( \dfrac{r}{5\mathrm{au}} \bigg)^{3/2}.
\label{eq:recovery_5AUref}
\end{equation}

Assuming outbursts occur regularly every $\Delta t_\mathrm{otb}$, we can estimate the fraction of time during which the dust population is out of local coagulation/fragmentation equilibrium as $t_\mathrm{rec}(r)/\Delta t_\mathrm{otb}$. Using $\Delta t_\mathrm{otb} = 10^4\mathrm{~yr}$, this fraction reaches $45\%$ for \texttt{B-por-m.s}. 

We can also find the position of the critical radius $r_\mathrm{crit}$, defined as the heliocentric distance outside which $t_\mathrm{rec}(r) > \Delta t_\mathrm{otb}$. If the outburst is sufficiently strong to push the water snowline outside the critical radius, an unrecovered annulus of width $r_\mathrm{SL}^\mathrm{otb} - r_\mathrm{crit}$ is formed, within which the dust distribution never reach the coagulation/fragmentation equilibrium (see Fig.~\ref{fig:cartoon_recovery}). The annulus presence and width is thus determined by the independent combination of outburst properties and coagulation physics. We find respectively $r_\mathrm{crit} = 23.2  \mathrm{~au}$ and $r_\mathrm{crit} = 14.6 \mathrm{~au}$ for \texttt{B-comp-resi} and \texttt{B-por-resi} respectively, and $r_\mathrm{crit} = 14.6 \mathrm{~au}$ and $r_\mathrm{crit} = 8.5 \mathrm{~au}$ for \texttt{B-comp-m.s} and \texttt{B-por-m.s}. These values are confirmed by the results of additional coagulation simulations not shown here. For our moderate-amplitude outburst, only aggregates obeying the many-seeds porous model would have an unrecovered annulus, located in between $r_\mathrm{crit} = 8.5 \mathrm{~au}$ and $r_\mathrm{SL}^\mathrm{otb} = 13\mathrm{~au}$. Other systems have been observed undergoing much stronger outbursts, like in V883 Ori where it has been suggested from HCO+ observations that the excited water snowline is located as far as $\approx 100\mathrm{~au}$ \citep[][]{leemker2021chemically}. In such a case, a considerable portion of the disc would not be expected to recover between repeated accretion outbursts.

\subsection{Dust emission as a past outburst tracer}
\label{sec:pastoutburst}

As mentioned in Sect.~\ref{sec:introduction}, one of our objective is to investigate whether the alteration of dust properties leaves a durable observational signature on discs. It would allow us to trace past outbursts, thus build a greater statistical estimate of sources undergoing such events to better understand their cause, strength and frequency. If most discs undergo repeated accretion outbursts during their lifetime, as suggested by the episodic accretion scenario \citep[][]{dunham2012resolving, audard2014episodic}, then the comprehension of such signatures would be of even greater importance for any protoplanetary discs observations. In this section, we will focus on the evolution of $\beta$, the spectral index of the dust opacity at millimetre wavelengths (see Sect.~\ref{sec:opacity_model}), as it is a quantity often accessible from ALMA observations.

We saw in Fig.~\ref{fig:beta} that the growth of compact aggregates is punctuated with a resonant amplification of $\beta$ around millimetre sizes. As the coagulation timescale increases with the heliocentric distance, the observation of a non-outbursting disc at a time $t$ should display a resonant peak at a specific radius $r$, propagating outward with time. The resonance being damped for porous aggregates, \citet[][]{kataoka2014opacity} predicted that the observation of such peak could infer the presence of compact growth in discs. In an outbursting system, however, an additional resonant signal could be present between the quiescent and excited snowline positions, due to the alteration of compact grains (Fig.~\ref{fig:beta}). This secondary signal would be visible during the outburst (for fast and intermediate adjustments, Fig.~\ref{fig:cartoon_recovery}) or after (slow adjustment). Given the relatively short duration of outbursts as compared to the coagulation timescale, it would most likely be present in the aftermath of the event, during the recovery phase. The secondary signal would be always visible in the zone B of a disc having an unrecovered annulus (Fig.~\ref{fig:cartoon_recovery}).

We therefore speculate that the observation of two resonant peaks at two radii of a protoplanetary disc could trace the occurrence of a past outburst event, in addition to tracking the compact structure of dust grains. The shape and height of the secondary peak could help predicting whether the compact aggregates follow the resilient or many-seeds model, thus using outbursting objects as laboratory to infer the behaviour of dust to sublimation. From predictions on the position of the quiescent snowline, the time passed since the outburst could be constrained, along with a lower estimate of the excited snowline position and the strength of the accretion outburst.

\begin{figure}
	\includegraphics[width=\columnwidth]{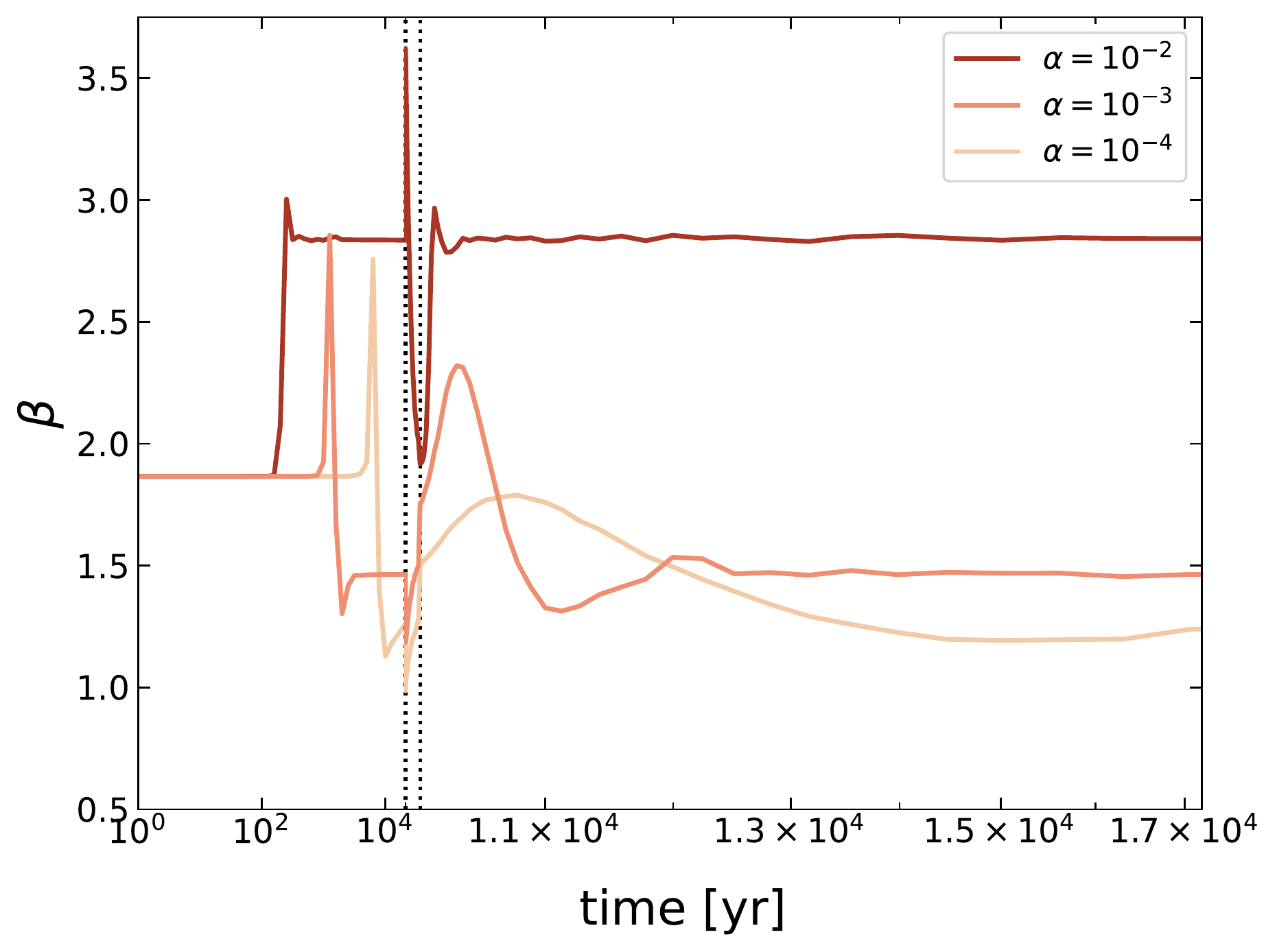}
    \caption{Identical to Fig.~\ref{fig:beta} for \texttt{B-comp-resi}, including two additional simulations performed respectively with stronger ($\alpha=10^{-2}$) and weaker ($\alpha=10^{-4}$) turbulence. Note that $t_\mathrm{otb}^\mathrm{start}=20000\mathrm{~yr}$, so that the simulation with low turbulence has the time to reach its coagulation/fragmentation equilibrium.}
    \label{fig:beta_turbulence}
\end{figure}

However, it is important to note a few factors that may impact that picture. First, even fairly moderate porosities are already enough to impact the appearance of the resonance peak. As shown for example in Fig.~3 of \citet[][]{miotello2022setting}, values corresponding to $\rho_\mathrm{int} \approx 10^{-1} \mathrm{~g~cm^{-3}}$ are sufficient to dampen the resonance peak. 
Second, even in the compact scenario several disc parameters may severely influence the particle size distribution and modify the temporal evolution of $\beta$ \citep[][]{birnstiel2011dust}. The turbulence, notably, can have a great impact, as higher turbulence leads to larger relative velocity and a lower maximum size in the coagulation/fragmentation equilibrium \citep[]{birnstiel2011dust}. We present on Fig.~\ref{fig:beta_turbulence} the temporal evolution of $\beta$ for the model \texttt{B-comp-resi} using different values of the turbulence parameter $\alpha$. For weaker turbulence ($\alpha=10^{-4}$), the size distribution reaches larger $a_\mathrm{max}$ leading to smaller $\beta$ (Fig.~\ref{fig:beta_vs_maxsize}). The post-outburst variations are weaker, and span on a longer duration ($\approx 5000 \mathrm{~yr}$). For stronger turbulence ($\alpha=10^{-2}$), $a_\mathrm{max}$ is located in the resonance size range, leading to a higher opacity index. It peaks at $\beta \approx 3.5$ when the composition changes, as the opacity index of the rocky mixture is higher (Fig.~\ref{fig:beta_vs_maxsize}). Similarly to \texttt{A-comp}, fragmentation leads to a mass loss of mm-sized aggregates and a sharp decrease in $\beta$, unlike the $\alpha=10^{-3}$ and $\alpha=10^{-4}$ cases. The post-outburst recovery is rapid, within $200\mathrm{~yr}$. We note that with settling effect included, the dust scale-height would also vary with the turbulence and the recovery timescale may differ from the results of our local model. In the assumption of Eq.~\ref{eq:recovery}, the recovery timescale would be independent of the turbulence strength.

\subsection{Implications on planetesimal formation}
\label{sec:planetesimal}

\begin{figure*}
    \hspace*{-1cm}                                                 
    \includegraphics[width=0.8\textwidth]{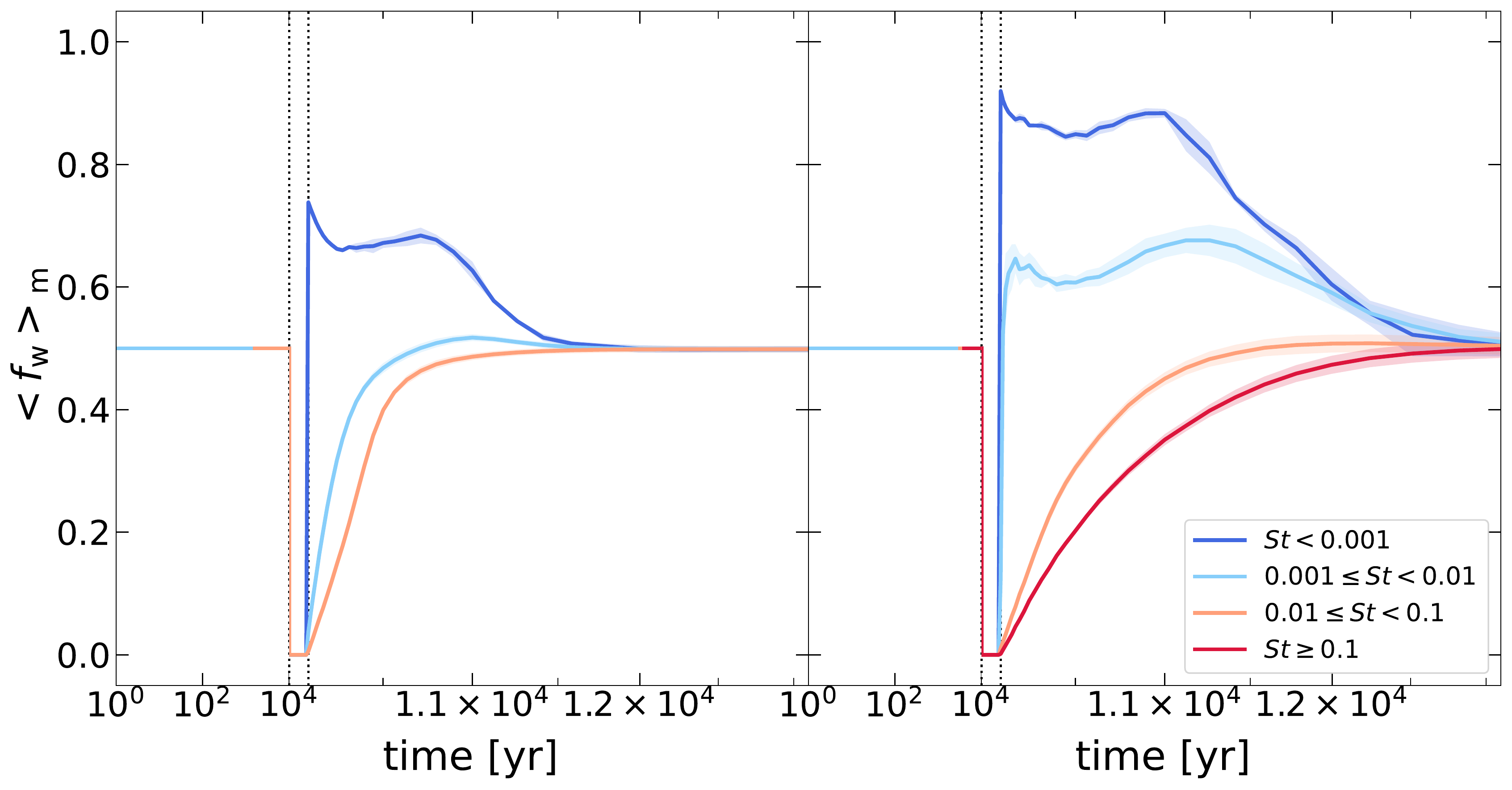}
    \caption{Mass-weighted average of the water fraction $f_\mathrm{w}$ for different Stokes number bins of compact (left) and porous (right) aggregates in the resilient model. The many-seeds is not represented as it displays negligible variation of the water fraction. The logarithmic scale is reset at $t_\mathrm{otb}^\mathrm{start}$ (see also Fig.~\ref{fig:max_size}). The statistical uncertainties are represented as shaded area and are obtained similarly to Fig.~\ref{fig:compact_BBseeds} and Fig.~\ref{fig:porous_ABBseeds}. }
    \label{fig:fw_massweighted}
\end{figure*}

Dust coagulation is the first step of planet formation, and according to the current consensus, is followed by the streaming instability (SI) to achieve the formation of km-sized planetesimals. Only sufficiently large pebbles are able to trigger the SI and form dense gravitationally unstable filaments \citep[][]{bai2010dynamics, drkazkowska2014can, li2021thresholds}, thus the properties of the largest pebbles (e.g. composition, structure) are inherited by the planetesimals. In this section, we will focus on the evolution of aggregates that could lead to the streaming instability, defined as having Stokes number $St \geq 0.01$ \citep[][]{li2021thresholds}, to infer the implications for the properties of planetesimals forming in outbursting systems. We will focus the discussion on zone B, as solids in the inner zone A do not reach the necessary conditions for SI (see Fig.~\ref{fig:max_size}).

In the resilient model, the occurrence of SI can be perturbed if fragmentation leads to an efficient mass loss of pebbles above $\mathrm{St}=0.01$, which is the case in our compact simulation at $5 \mathrm{~au}$. Concerning the water fraction, both compact and porous pebbles within zone B are totally deprecated in water ice for the duration of the outburst. If the streaming instability occurs during that period, it will form ice-free planetesimals. After the outburst, pebbles remain poorer in water for a more or less extended time depending on the efficiency of collisional mixing with ice-rich grains. We can see in Fig.~\ref{fig:fw_massweighted} that it takes at $5\mathrm{~au}$ about $500\mathrm{~yr}$ for compact pebbles to recover most of their initial water content, and between $1000$ and $2000\mathrm{~yr}$ for porous pebbles depending on their Stokes number. Planetesimals forming during that mixing period will be poorer in water ice than in a non-outbursting system. The collision rates decreasing with heliocentric distance, these timescales could be even greater especially during stronger outbursts \citep[e.g. $r_\mathrm{SL}^\mathrm{otb} \approx 100\mathrm{~au}$ in V883 Ori][]{leemker2021chemically} favoring the arising of an unrecovered annulus (Fig.~\ref{fig:cartoon_recovery}) in which planetesimals would form poorer in water ice at any time of the disc lifetime. Concerning the planetesimals structure, \citet[][]{visser2021radial} argued that during their formation through SI, pebbles collapse towards the core from the largest Stokes number to the smallest. As seen in Fig.~\ref{fig:fw_massweighted}, we can expect to find a compositional radial gradient within planetesimals, with a more rocky core and most of the -- already lowered -- water ice located in the loose upper layers, more sensible to disruptions by e.g. close encounter with the host star \citep[][]{visser2021radial}. Concerning the porosity, the internal density of pebbles can be inherited by the planetesimals in case of a gentle cloud collapse \citep[typically for planetesimals $<100\mathrm{~km}$, see ][]{jansson2014formation}. However, we do not find the accretion outburst to have a notable impact on the porosity of the largest aggregates participating in the streaming instability, remaining around $\rho_\mathrm{int} \approx 10^{-3} \mathrm{~~g~cm^{-3}}$ (see Fig.~\ref{fig:porous_ABBseeds}).

In the many-seeds model, aggregates fall apart to monomer size, completely emptying the reservoir of pebbles in the entire zone B, and halting the formation of planetesimals on scales $\geq 10 \mathrm{~au}$ depending on the outburst strength. Depending on $t_\mathrm{rec}$ (Sect.~\ref{sec:postoutburstrecovery}), it may take several thousands of years for aggregates to grow back to their pre-outburst state and to be available for forming planetesimals through SI\footnote{As SI does not need the population to fully recover to its collisional equilibrium, but just grow past $\mathrm{St}=0.01$, the time is slightly smaller than $t_\mathrm{rec}$ (up to a factor $\approx 2$ in our porous simulations).}. Assuming a constant outburst rate, the formation of planetesimal at a radius $r$ in zone B will be halted for a fraction $t_\mathrm{rec}(r)/\Delta t_\mathrm{otb}$ of the disc lifetime, representing respectively $20\%$ and $45\%$ for the compact and porous growth at $5 \mathrm{~au}$. In the case of a slow adjustment (see Fig.~\ref{fig:cartoon_recovery}), the formation of planetesimals in the unrecovered annulus between $r_\mathrm{crit}$ and $r_\mathrm{SL}^\mathrm{otb}$ may be completely inhibited for the entirety of the disc lifetime.

\subsection{Limitation of the local approach}
\label{sec:radialdriftlimitation}
The clear limitation of our Monte Carlo coagulation code lies in its local approach. Even though we consider the radial drift for the calculation of the relative velocity, we cannot take into account that solids may be removed from their environment due to efficient drift, notably around $St \approx 1$. In particular, if $t_\mathrm{drift} < t_\mathrm{grow}$ for sizes below the fragmentation limit, then the growth would be halted by the radial drift, i.e. the radial drift barrier. 

We estimate the growth timescale at a given size with $t_\mathrm{grow} = a/(da/dt)$ and the corresponding drift timescale as the orbital radius divided by the radial drift velocity. In the inner disc, the condition
$t_\mathrm{drift} > t_\mathrm{grow}$ is largely satisfied for all sizes. In location B, calculations show that $t_\mathrm{drift}>10^5\mathrm{~yr}$, i.e. the length of the simulation, for almost all sizes. The drift timescale being typically larger than: 1) the time needed for the population to reach its coagulation/fragmentation equilibrium, 2) the outburst duration, and 3) the recovery timescale, we do not expect radial drift to significantly alter our findings. However, for the very largest particles close to the fragmentation barrier - with sizes ${>}0.2\mathrm{~cm}$ and ${>}1400\mathrm{~cm}$ respectively for compact and porous aggregates - the drift timescale can be as short as $t_\mathrm{drift} \lesssim 10^4\mathrm{~yr}$. While this is still long compared to the outburst duration and recovery time, considerable radial drift may take place between outbursts. Including the effects of radial transport will be the focus of future work.

\subsection{Conservation of the total water mass}
\label{sec:conserving_water}
Since we are considering a closed volume of dust and gas in a disc, the total mass of rock and water (in ice or vapour) should be conserved in time. The total rock mass conservation is ensured by our definition of the swarms, as the mass of each swarm $M_\mathrm{swm}$ is kept constant through the growth by adjusting the number of particles $N_i$ in that swarm. However, because of how the superparticle approach only updates the $i$-th particle involved in the collision, fluctuations in the total water mass may appear if the colliding pair ($i$, $j$) has dissimilar water fractions, which happens for example in location B after the outburst. After the water content is re-distributed through the population by collisional mixing, the fluctuations disappear and the total water mass stabilises. Simulations used in this paper, with $n=10^4$ superparticles, displayed fluctuations of the total water mass overall averaging below $1\%$. For \texttt{B-por-resi}, which displayed the broadest water distribution (Fig.~\ref{fig:porous_ABBseeds}), the fluctuations for an individual run can be as high as $5.7\%$, but averaging over the three independent runs leads to $0.3\%$. With such values, the variation of the total water mass does not have a noticeable impact on our results, but we note that using the Monte Carlo approach originally proposed by \citet[][]{ormel2007dust} could remove such statistical fluctuations, as the properties of both colliding particles are updated. 

\subsection{Other ice species}

Throughout this manuscript, we focused only on the impact of water ice. We note that including other volatile species and snowlines (e.g. CO, $\mathrm{CO}_2$) could be an interesting direction for future work. However, there is still a large parameter space to explore by laboratory experiments concerning the impact of multiple ice species on the collisional properties of dust grains, and whether the many-seeds response can be extended to the sublimation of other abundant ices. We also note that dust in high temperature environments ($T > 1200 \mathrm{~K}$) is thought to become more sticky \citep[][]{pillich2021drifting}. Accretion outbursts could provide the necessary temperature to lead to boosted growth in a more extended fractions of the inner disc. Whether that may offer a pathway for the formation of terrestrial planets could be a key aspect to explore.

\section{Summary and outlook}
\label{sec:conclusions}

We have developed a local coagulation model based on the superparticle approach \citep[][]{zsom2008representative} to simulate dust growth/fragmentation in a disc undergoing an FUor-type accretion outburst (Sect.~\ref{sec:simulationmethod}). We followed the evolution of grain properties, and considered multiple structural designs for the aggregation and response to sublimation (summarised in Fig.~\ref{fig:cartoon_dustmodel}). We applied our model at two disc locations to explore the impacts of the outburst with and without compositional changes (Sect.~\ref{sec:discmodel}). Coagulation results (Sect.~\ref{sec:results}) were then converted into absorption opacity to investigate whether the alteration of dust properties has implications for the observation of protoplanetary discs (Sect.~\ref{sec:opacity_model}). Our main findings are summarised as follows:

\begin{enumerate}[label={\arabic*.}]
    
    \item The accretion outburst affects the size distribution in the entire disc and for all dust models (e.g. Sect.~\ref{sec:result_compact} and Sect.~\ref{sec:result_porous}). The most dramatic size alteration occurs in Zone B, and when particles fall apart upon water sublimation (i.e. many-seeds model). If aggregates survive sublimation (i.e. resilient model), the size reduction is driven by fragmentation and depends on the time required by pebbles to recover their initial water content. In zone A and C, the size alteration is smaller as the change in temperature is not accompanied by modifications of dust properties (see Sect.~\ref{sec:results_compact_A}).
    
    \item Only solids in zone A adjust to the new collisional equilibrium within the end of the outburst (Sect.~\ref{sec:maximumsize}). In zone B, the size distribution takes longer to adjust and its peak is not well characterized by a theoretical fragmentation limit (Fig.~\ref{fig:max_size}). In the many-seeds model aggregates are generally smaller than the theoretical maximum size $a_\mathrm{max}^\mathrm{rock}$. In the resilient scenario aggregates will instead be too large for their ice-free composition (see slow adjustment in Fig.~\ref{fig:cartoon_recovery}). The latter may offer an explanation for the observation of large dry pebbles in FU Ori by \citet[][]{liu2021millimeter}.

    \item Re-condensation leads to an heterogeneous distribution of water, preferentially depositing ice on small grains dominating the total surface area  (Sect.~\ref{sec:result_compact} and Sect.~\ref{sec:result_porous}), which highly diverges from the constant $f_\mathrm{w}$ expected in non-outbursting systems. The water fraction and internal density distributions are mostly affected if aggregates have a broad size range at the time of re-condensation. This is the case in the resilient model, and could be the case in the many-seeds model if e.g. the outburst is longer ($\tau_\mathrm{otb} > 100\mathrm{~yr}$) or the turbulence stronger ($\alpha > 10^{-3}$) than in our simulations (Sect.~\ref{sec:maximumsize}). The time needed to recover the pre-outburst water distribution depends on the efficiency of collisional mixing between pebbles and dust grains, reaching more than $1000 \mathrm{~yr}$ in the porous resilient model at $5 \mathrm{~au}$ (Fig.~\ref{fig:fw_massweighted}).
    
    \item After the accretion outburst, the population returns to the initial equilibrium on a timescale that depends on the outburst duration, coagulation physics, aggregation model, and the response to sublimation (Sect.~\ref{sec:postoutburstrecovery}). In our simulations at $5\mathrm{~au}$, it takes up to $4500 \mathrm{~yr}$ for porous many-seeds aggregates (Fig.~\ref{fig:max_size}). Depending on how the recovery timescale compares to the outburst rate, there may be portions of the disc where solids never reach coagulation/fragmentation equilibrium (i.e. unrecovered annulus, see Fig.~\ref{fig:cartoon_recovery}).

    \item The changes in size distribution and ice content together result in a complex response in the absorption opacity (Fig.~\ref{fig:totalopacity}) also visible at millimetre wavelengths through the opacity index $\beta$ (Fig.~\ref{fig:beta}). Dust emissions behave quite differently whether aggregates have a compact or a porous structure \citep[][]{kataoka2014opacity}. At millimetre wavelengths, emissions are optically thin at $5\mathrm{~au}$ (Fig.~\ref{fig:opt_depth}).

    \item If dust particles are compact, the opacity index $\beta$ would be a good indicator of their alteration by the outburst. In our simulations, the recovery of aggregates leads to a sharp increase of $\beta$ after the event, reaching $\beta \approx 2.9$ after $1200\mathrm{~yr}$ in the many-seeds case. This observational feature may provide a way to track past accretion outbursts in protoplanetary discs, and improve our statistical sample of such events. In addition, the distinct profiles associated to the resilient and many-seeds models could allow to determine how aggregates actually respond upon sublimation, making outbursting objects important laboratories for exploring the structure of dust particles (Sect.~\ref{sec:pastoutburst}).

    \item The formation of planetesimals is impacted by the outburst (Sect.~\ref{sec:planetesimal}). In the resilient case, efficient fragmentation leads to a mass loss of large pebbles which can lower the chance to trigger planetesimal formation through the streaming instability. If they do form, their properties will be set by the altered properties of pebbles, i.e. ice-free during the outburst, and ice-poor after, for a duration dependant of the efficiency of collisional mixing with ice-rich grains. It would additionally lead to a composition radial gradient in their structure (Fig.~\ref{fig:fw_massweighted}). In the many-seeds case, their formation through the streaming instability is inhibited for the time required to re-grow large pebbles (up to $4500\mathrm{~yr}$ in the porous model).

\end{enumerate}

In summary, our simulations have demonstrated how FUor-type accretion outbursts can alter the collisional evolution of dust and ice in protoplanetary disc midplanes, leading to changes in e.g., the ice distribution and maximum size that persist long after the outburst has faded. Given that most systems are thought to experience such frequent outbursts during their evolution, as suggested in the episodic accretion scenario \citep[][]{dunham2012resolving, audard2014episodic}, we stress that considering their impact on dust evolution is a key aspect to further understand the structure of protoplanetary discs and the process of planet formation. Investigating further the recovery front in the radial profile of the disc and comparing with observations will be the focus of follow-up works.

\section*{Acknowledgements}

We are grateful to the anonymous reviewer for their thorough and insightful comments which helped improve the manuscript. We thank Enrique Mac{\'\i}as for useful discussions regarding interpretations of ALMA observations and the opacity index, and David J. Simon for helpful comments on the design of the schematics present in the manuscript. This project has made use of the package \texttt{DSHARP-OPAC} \citep[][]{birnstiel2018disk}, along with the following Python packages: \texttt{MATPLOTLIB} \citep[][]{hunter2007matplotlib}, \texttt{NUMPY} \citep[][]{harris2020array}, and \texttt{PANDAS} \citep[][]{mckinney2010data}.

%%%%%%%%%%%%%%%%%%%%%%%%%%%%%%%%%%%%%%%%%%%%%%%%%%
\section*{Data Availability}

The particles properties generated with our coagulation simulations used for this paper will be shared upon reasonable request. The \texttt{DSHARP-OPAC}\footnote{\url{https://github.com/birnstiel/dsharp_opac}} package developed by \citet[][]{birnstiel2018disk} is publicly available.

\bibliographystyle{mnras}
\bibliography{paper1}

%%%%%%%%%%%%%%%%% APPENDICES %%%%%%%%%%%%%%%%%%%%%

\appendix
\section{Assessing the position of the collisional equilibrium}
\label{sec:appendixA}

\begin{figure*}
	\includegraphics[width=\textwidth]{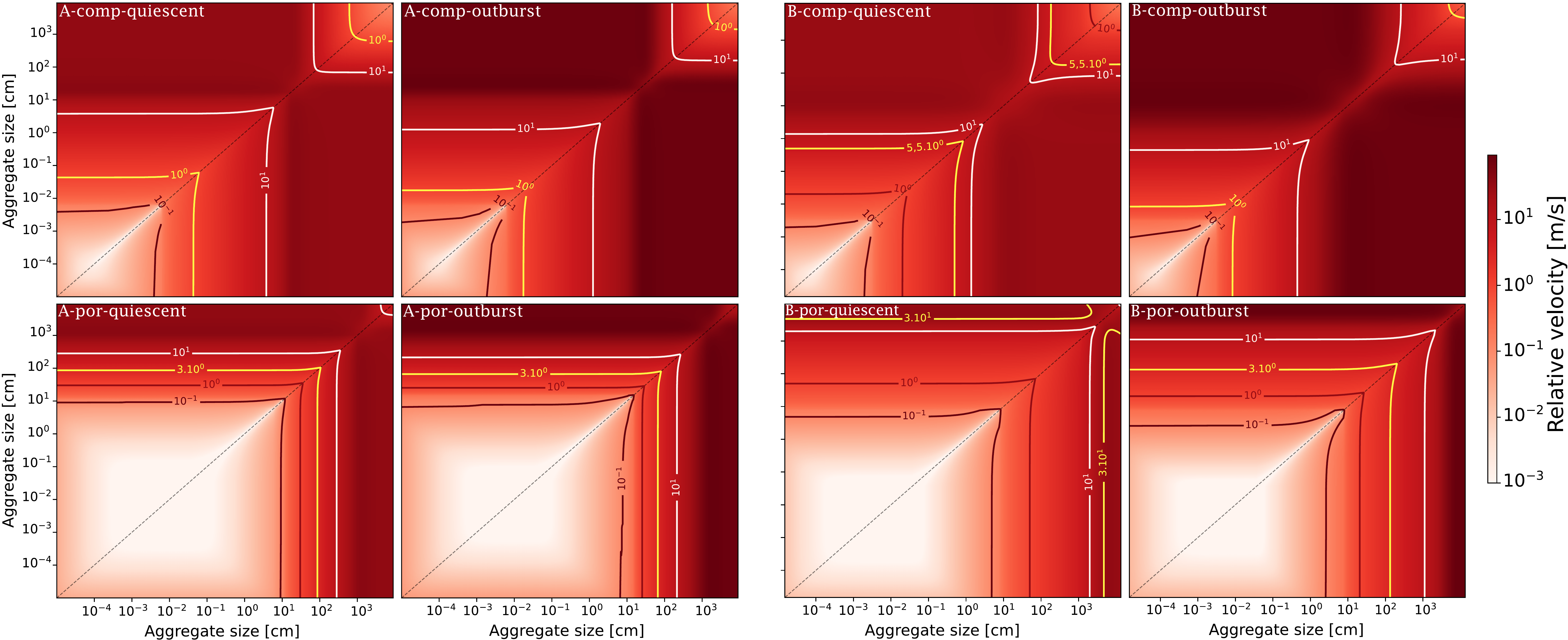}
    \caption{Total relative velocity between aggregates with varying size in location A (left) and B (right), accounting for the contributions of the Brownian motion, turbulence, radial and azimuthal drifts. The dashed diagonal line represents equally-sized particles, and the contours indicate a few values of the relative velocity ($0.1$, $1$, and $10 \mathrm{~m~s^{-1}}$) in addition to $v_\mathrm{f}$ outlined in yellow. The relative velocity is calculated during the quiescent and outburst phase. Compact aggregates are on the top panels, and porous ones on the bottom.}
    \label{fig:velocity_A_B}
\end{figure*}

The relative velocity between dust particles is related to the local disc conditions and to their respective aerodynamical behaviour. Typically, it increases with particle size. We compute the total relative velocity for varying collider size in Fig.~\ref{fig:velocity_A_B} for location A and B. As seen in Sect.~\ref{sec:dustdynamics}, we include the contributions of the Brownian motion, turbulence, and radial and azimuthal drifts. The left (resp. right) panels represent the quiescent (resp. outburst) disc conditions. The top (resp. bottom) panels show the compact (resp. porous) aggregation. The yellow contours indicate the location of the corresponding fragmentation limit $v_\mathrm{f}$ (Sect.~\ref{sec:aggregation}). 

We determine the approximate value of $\langle a \rangle_\mathrm{m}$ at the coagulation/fragmentation equilibrium by solving for the size at which equal aggregates (dashed diagonal line) collide at $v_\mathrm{f}-\delta v_\mathrm{f}$, where $\delta v_\mathrm{f}$ represents the width of the transition regime between sticking and fragmentation (Sect.~\ref{sec:superparticleapproach}). In fact, above that relative velocity, the collisions of equal aggregates begin to outcome on mass loss. If the system is given enough time to evolve, this size should represent an estimate of the upper region of the size distribution. As the largest aggregates generally dominate the mass of the system, it is thus a good estimate of the mass-weighted average size $\langle a \rangle_\mathrm{m}$. Note that in the icy porous case, equal aggregates do not reach $v_\mathrm{f}$, hence fragmentation should only occur in the transition regime or between aggregates with different sizes satisfying the condition $R_\mathrm{m} \geq 0.01$ (Sect.~\ref{sec:compactaggregation}). The size estimates for the quiescent equilibrium and outburst equilibrium are shown in Table \ref{tab:peak_size_at_equilibrium} for the different models. The response to sublimation having no influence on the fragmentation velocity, the values are identical for these models.

\begin{table}
	\centering
	\caption{Estimates of the mass-weighted average size at the collisional equilibrium for the different models in quiescent and outburst conditions.}
	\label{tab:peak_size_at_equilibrium}
	\begin{tabular}{lccc} % four columns, alignment for each
		\hline
		Model & Quiescent $\rightleftharpoons$ ($\mathrm{cm}$)  & Outburst $\rightleftharpoons$ ($\mathrm{cm}$) \\
		\hline
		\texttt{A-comp} & $0.013$ & $0.005$  \\
		\texttt{A-por} & $12.86$ & $9.56$  \\
		\texttt{B-comp-resi} \& \texttt{B-comp-m.s} & $0.178$ & $0.002$  \\
		\texttt{B-por-resi} \& \texttt{B-por-m.s} & $1042.43$ & $43.9$  \\
		\hline
	\end{tabular}
\end{table}

%%%%%%%%%%%%%%%%%%%%%%%%%%%%%%%%%%%%%%%%%%%%%%%%%%

% Don't change these lines
\bsp	% typesetting comment
\label{lastpage}
\end{document}